\documentclass[12pt]{article}
\RequirePackage[OT1]{fontenc}
\RequirePackage{amsthm,amsmath}
\RequirePackage[authoryear]{natbib}

\usepackage{color, dsfont}
\usepackage{url}
\usepackage{amsmath,amsfonts,amsthm,amssymb,amsbsy,bbm}
\usepackage{multirow}
\usepackage{listings}
\usepackage[utf8]{inputenc}
\usepackage{graphicx}
\usepackage{array}
\usepackage{tablefootnote}
\usepackage[dvipsnames]{xcolor}
\usepackage{array,booktabs,multirow,cleveref}
\usepackage{diagbox}

\newtheorem{algorithm}{Algorithm}
\newtheorem{theorem}{Theorem}
\newtheorem{lemma}{Lemma}
\newtheorem{corollary}{Corollary}

\theoremstyle{definition}

\newtheorem{example}{Example}
\newtheorem{remark}{Remark}

\usepackage{adjustbox}
\usepackage{float}
\usepackage{makecell}
\usepackage{pgfplots}
\usepgfplotslibrary{fillbetween}

\begin{document}

\title{\large\bf Constrained D-optimal Design for Paid Research Study}
\author{Yifei Huang$^{1}$, Liping Tong$^{2}$ and Jie Yang$^{1}$\\
	$^1$University of Illinois at Chicago~and $^2$Advocate Aurora Health
}

\maketitle
	
\begin{abstract}
We consider constrained sampling problems in paid research studies or clinical trials. When qualified volunteers are more than the budget allowed, we recommend a D-optimal sampling strategy based on the optimal design theory and develop a constrained lift-one algorithm to find the optimal allocation. Unlike the literature which mainly deals with linear models, our solution solves the constrained sampling problem under fairly general statistical models, including generalized linear models and multinomial logistic models, and with more general constraints. We justify theoretically the optimality of our sampling strategy and show by simulation studies and real-world examples the advantages over simple random sampling and proportionally stratified sampling strategies.  
\end{abstract}

{\it Key words and phrases:}
Constrained sampling, D-optimal design, Generalized linear model, Lift-one algorithm, Multinomial logistic model
		
\def\thefigure{\arabic{figure}}
\def\thetable{\arabic{table}}

\numberwithin{equation}{section}

\section{Introduction}
\label{sec:intro}

We consider a constrained sampling problem frequently arising in paid research studies or clinical trials, especially when recruiting volunteers via the Internet or emails, which could gather attention widely and quickly. For example, some investigators plan to conduct a research study to evaluate the effect of a new treatment on anxiety. Besides the treatment cost, the investigators also need to prepare certain compensation for participants' time. Due to limited funding, the investigators could only support up to $n$ participants while there are $N > n$ eligible volunteers. The question is how they select $n$ participants out of $N$ to evaluate the treatment effect most accurately.
Noted that the goal of the sampling problem in this paper is not the mean of response but the treatment effect or regression coefficients of an underlying statistical model.

A straightforward approach is to use the {\it simple random sampling without replacement} (SRSWOR, see, for example, Chapter~2 of \cite{lohr2019sampling}), which randomly chooses an index set $1\leq i_1 < i_2 < \cdots < i_n \leq N$ such that each index set of $n$ distinct subjects has the equal chance $n!(N-n)!/N!$ to be chosen. This can be applied if the investigators know nothing about the volunteers except contact information, or the covariate information provided by the volunteers does not seem relevant to the treatment effect. 

A more common practice is, however, that the investigators collect some covariates information, such as gender and age, which may play some roles in the treatment effect. Suppose there are $d$ covariates and $m$ distinct combinations of covariates under consideration. For example, $d=2$ covariates, gender (male or female), and age (18$\sim$25, 26$\sim$64, 65 and above), lead to $m=6$ possible categories (combinations of covariates) of eligible volunteers, known as {\it strata} in the sampling theory (see, for example, Chapter~3 in \cite{lohr2019sampling}). Suppose the frequencies of volunteers in the $m$ categories are $N_1, \ldots, N_m$, respectively. The question is how we determine $n_i \leq N_i$ such that $n=\sum_{i=1}^m n_i$, known as the {\it allocation} of subjects to the categories or strata. Once an allocation $(n_1, \ldots, n_m)$ is determined, $n_i$ subjects will be chosen randomly from the $i$th category or stratum for each $i$, known as a {\it stratified} (random) sampling. A commonly used stratified sampler chooses $n_i \propto N_i$, known as the {\it proportionally stratified} sampler, which is expected to produce a more accurate estimate for the mean response than SRSWOR (Section~3.4.1 in \cite{lohr2019sampling}). 

Nevertheless, from an optimal design point of view \citep{fedorov1972, silvey1980, pukelsheim1993, atkinson2007, fedorov2014}, we want to find $n_i$'s such that the treatment effects or regression coefficients can be estimated most accurately. When no prior knowledge about the regression model is available, a {\it uniform} allocation, which assigns roughly the same number of subjects to each category, is commonly used in the practice of experimental design (see, for example, \cite{yang2012optimal}). For the sampling problem under constraints $n_i\leq N_i$, $i=1, \ldots, m$, we recommend (constrained) {\it uniformly stratified} sampler, which chooses $n_i = \min\{k, N_i\}$ or $\min\{k, N_i\}+1$ with $k$ satisfying $\sum_{i=1}^m \min\{k, N_i\} \leq n < \sum_{i=1}^m \min\{k+1, N_i\}$ (see Section~\ref{sec:doptimal}). 
 
If the investigators have some information about the regression coefficients from some pilot study or prior research, we propose optimal stratified samplers based on the optimal design theory, which minimizes the variances of the estimated regression coefficients instead of the estimated population mean. According to different optimality criteria used \citep{fedorov1972, atkinson2007, stufken2012, fedorov2014}, we call the corresponding sampler D-optimal sampler, A-optimal sampler, etc. In this paper, we focus on D-optimality, which is the most frequently used \citep{atkinson1999, atkinson2007,ytm2016}.

In the statistical literature, optimal designs under constraints were considered mainly for linear models with the information ${\mathbf F}_i = {\mathbf h}({\mathbf x}_i) {\mathbf h}({\mathbf x}_i)^T$ obtained at the $i$th experimental setting ${\mathbf x}_i$ (see Section~\ref{sec:constrained_optimal_allocation}), where ${\mathbf h}({\mathbf x}) = (h_1({\mathbf x}), \ldots, h_p({\mathbf x}))^T$ are known predictor functions (see \cite{elfving1952optimum, lee1988constrained, cook1995invited, fedorov2014} and references therein). Among them, \cite{wynn1977minimax, wynn1977optimum, wynn1982optimum} connected finite population sampling with optimal designs under constraints $n_i \leq N_i$ (Example~\ref{ex:2*3}); and \cite{welch1982branch}, \cite{fedorov1989optimal}, \cite{pronzato2004minimax, pronzato2006sequential} developed algorithms searching for ``optimum submeasures'' or ``optimum bounded designs''. In this paper, ${\mathbf F}_i$ could be much more complicated and depend on unknown parameters $\boldsymbol\theta$, such as $\nu\{{\mathbf h}({\mathbf x}_i)^T \boldsymbol\theta\} {\mathbf h}({\mathbf x}_i) {\mathbf h}({\mathbf x}_i)^T$ for generalized linear models (Section~\ref{sec:doptimal}) or ${\mathbf X}_i^T {\mathbf U}_i {\mathbf X}_i$ for multinomial logistic models (Section~\ref{sec:categorical}).
More than that, the design problem discussed here is under more general constraints including but not limited to $n_i\leq N_i$, $n_i + n_j \leq N_{ij}$ (Example~\ref{ex:trauma}), $4n_i \geq n_j$ (Example~\ref{ex:counter}), where $N_{ij}$ is a pre-determined upper bound for the $i$th and $j$th categories in total. 

For unconstrained optimal design problems, many numerical algorithms have been proposed using directional derivatives \citep{wynn1970, fedorov1972, atkinson2014, fedorov2014}. Among them, the lift-one algorithm \citep{ymm2016, ym2015, ytm2016, bu2020} breaks the problem into univariate optimizations, utilizes analytic solutions whenever possible, reduces unnecessary weights to exact zeros, and works the same well for both local D-optimality and EW D-optimality. A comprehensive numerical study by  \citeauthor{ymm2016} (2016, Table~2) shows that the lift-one algorithm is more efficient than many other commonly used optimization
techniques and design algorithms for similar purposes. Unfortunately, it does not work in general for our constrained optimal design problems (see Subsection~\ref{sec:counterexample}). 
The sequential number-theoretic optimization (SNTO) algorithm \citep{fang1993number, gong1999improved, gao2022high} may provide a solution for our constrained optimization problems, which is, however, typically not as efficient as optimization algorithms based on directional derivatives when the objective function is differentiable and unimodal \citep{FangKai-Tai1994SAoN}. 

In this paper, we develop a new algorithm, called the {\it constrained lift-one algorithm}, to find optimal allocations under fairly general constraints and statistical models. While keeping the high efficiency of the original lift-one algorithm, the proposed algorithm will check the optimality of the converged allocation and utilize linear programming for adjusting the searching direction when needed. We provide theoretical justifications for the optimality of the allocation found by the proposed algorithm (Section~\ref{sec:constrained_lift_one}). Our simulation studies with generalized linear models (Section~\ref{sec:doptimal}) and a real data example with multinomial logistic models (Section~\ref{sec:categorical}) show that uniformly stratified sampler usually works better than SRSWOR and proportionally stratified sampler, and our designer's choice, (locally) D-optimal and EW D-optimal samplers can significantly improve the efficiency further when some information about the regression coefficients is available.

\section{Constrained D-optimal Allocation}\label{sec:constrained_optimal_allocation}

In general, we consider an experiment with $m\geq 2$ pre-determined experimental settings or level combinations ${\mathbf x}_i = (x_{i1}, \ldots, x_{id})^T \in \mathbb{R}^d$ of $d$ covariates. Suppose we allocate $n_i \geq 0$ subjects to the $i$th experimental setting ${\mathbf x}_i$, and the responses are independent and follow a parametric model $M({\mathbf x}_i; \boldsymbol\theta)$ with unknown parameters $\boldsymbol\theta \in \boldsymbol\Theta \subseteq \mathbb{R}^p$, $p\geq 2$. Under regularity conditions, the Fisher information matrix of the sample can be written as ${\mathbf F} = \sum_{i=1}^m n_i{\mathbf F}_i \in \mathbb{R}^{p\times p}$, where ${\mathbf F}_i$ corresponds to the Fisher information at ${\mathbf x}_i$ and is a positive semi-definite matrix (see, for example, Section~1.5 in \cite{fedorov2014} and references therein).

In design theory, ${\mathbf n} = (n_1, \ldots, n_m)^T$ is called an {\it exact} allocation of $n=\sum_{i=1}^m n_i$ experimental units, while ${\mathbf w} = (w_1, \ldots, w_m)^T = (n_1/n, \ldots, n_m/n)^T$ is called an {\it approximate} allocation, which is easier to be dealt with theoretically. The constrained D-optimal design problem considered in this paper is to find the approximate allocation ${\mathbf w}$, which maximizes $|{\mathbf F}|$, the determinant of $\mathbf F$, on a collection of feasible approximate allocations $S \subset S_0 := \{(w_1, \ldots, w_m)^T$ $\in$ $\mathbb{R}^m \mid w_i \geq 0, i=1, \ldots, m; \sum_{i=1}^m w_i = 1\}$. We assume that $S$ is either a closed convex set itself or a finite (overlapped or disjoint) union of closed convex sets. If $S = \cup_{k=1}^K S_k$, where $S_k$'s are all closed convex subsets of $S_0$, we can always find an optimal allocation for each $S_k$ and then pick up the best one among the optimal allocations. Therefore, in theory, we only need to solve the case when $S$ itself is closed and convex.

\begin{example}\label{ex:2*3}{\rm
Consider a paid research study with $N=500$ eligible volunteers. Suppose $d=2$ covariates, gender ($x_{i1} = 0$ for female and $1$ for male) and age group ($x_{i2} = 0$ for $18\sim 25$, $1$ for $26\sim 64$, and $2$ for $65$ or above), are important factors. There are $m=6$ categories with ${\mathbf x}_i = (x_{i1}, x_{i2})^T$ corresponds to $(0,0), (0,1), (0,2), (1,0), (1,1), (1,2)$, respectively. Suppose the numbers of volunteers $N_i$ in the $i$th category are $50, 40, 10, 200, 150, 50$, respectively. The budget can only support up to $n=200$ participants who will be under the same treatment. Their responses that will be recorded are binary, $0$ for no effect, and $1$ for effective. The goal is to study how the effective rate changes along with gender and age group. The collection of feasible approximate allocations is $S = \{(w_1, \ldots, w_6)^T \in S_0 \mid n w_i \leq N_i, i=1, \ldots, 6\}$, which is closed and convex.
\hfill{$\Box$}
}\end{example}

\begin{example}\label{ex:trauma}{\rm
\citeauthor{chuang1997} (1997, Table~V) provided a dataset of $N=802$ trauma patients, stratified according to the trauma severity at the time of study entry with $392$ mild and $410$ moderate/severe patients enrolled. The study involved four treatment groups determined by {\tt dose} level, $x_{i2} = 1$ ({\tt Placebo}), $2$ ({\tt Low dose}), $3$ ({\tt Medium dose}), and $4$ ({\tt High dose}). Combining with {\tt severity} grade ($x_{i1} = 0$ for {\tt mild} or $1$ for {\tt moderate}/{\tt severe}), there are $m=8$ distinct experimental settings with $(x_{i1}, x_{i2}) = (0,1), (0,2), (0,3), (0,4), (1,1), (1,2), (1,3), (1,4)$, respectively. The responses belong to five ordered categories, {\tt Death} ($1$), {\tt Vegetative state} ($2$), {\tt Major disability} ($3$), {\tt Minor disability} ($4$) and {\tt Good} {\tt recovery} ($5$), known as the Glasgow Outcome Scale \citep{jennett1975}. Suppose due to a limited budget, only $n=600$ participants could be supported, the collection of feasible approximate allocations is $S=\{(w_1, \ldots, w_8)^T \in S_0\mid n(w_1+w_2+w_3+w_4) \leq 392, n(w_5+w_6+w_7+w_8) \leq 410\}$, which is closed and convex.
\hfill{$\Box$}
}\end{example}

In this paper, we adopt the D-optimality, which is to maximize the objective function $f({\mathbf w}) = \left|\sum_{i=1}^m w_i {\mathbf F}_i\right|$, ${\mathbf w} \in S$. To avoid trivial cases, we assume that $f({\mathbf w}) > 0$ for some ${\mathbf w} \in S$. For statistical models under our consideration, such as typical generalized linear models (see Section~\ref{sec:doptimal}) and multinomial logistic models (see Section~\ref{sec:categorical}), ${\rm rank}({\mathbf F}_i) < p$ for each ${\mathbf x}_i \in {\cal X}$, the collection of all feasible design points, known as the {\it design space}. Although positive definite ${\mathbf F}_i$ exists for some special multinomial logistic models, it is uncommon and out of the scope of this paper.

\begin{lemma}\label{lem:w_i_less_than_1}
Suppose ${\rm rank}({\mathbf F}_i) < p$ for each $i$. If $f({\mathbf w}) > 0$ for some ${\mathbf w} = (w_1, \ldots, w_m)^T \in S$, then $0\leq w_i < 1$ for each $i$.
\end{lemma}

\begin{theorem}\label{thm:existence_general}
Suppose $S\subseteq S_0$ is closed and $f({\mathbf w}) > 0$ for some ${\mathbf w} \in S$. Then $f({\mathbf w})$ is an order-$p$ homogeneous polynomial of $w_1, \ldots, w_m$, and a D-optimal allocation ${\mathbf w}_*$ that maximizes $f({\mathbf w})$ on $S$ must exist.
\end{theorem}

Lemma~\ref{lem:w_i_less_than_1} and Theorem~\ref{thm:existence_general} confirm the existence of the constrained D-optimal allocation. Their proofs, as well as proofs of all other lemmas and theorems, are relegated to Section~S6 in the Supplementary Material.

For nonlinear models \citep{fedorov2014}, generalized linear models \citep{ym2015}, or multinomial logistic models \citep{bu2020}, ${\mathbf F}$ depends on the unknown parameters $\boldsymbol\theta$. We need an assumed $\boldsymbol\theta$ to obtain a D-optimal allocation, known as a {\it locally} D-optimal allocation \citep{chernoff1953}. When the investigators only have a rough idea about the parameters, with an assumed prior distribution on $\boldsymbol\Theta$, the parameter space, Bayesian D-optimality \citep{chaloner1995} maximizes $E(\log|{\mathbf F}|)$ and provides a more robust allocation. To overcome its computational intensity, an alternative solution, the EW D-optimality \citep{atkinson2007, ymm2016}, which maximizes $\log |E({\mathbf F})|$ or $|E({\mathbf F})|$, was recommended by \cite{ymm2016, ytm2016} and \cite{bu2020} for various models.  In this paper, we focus on local D-optimality and EW D-optimality.

\section{Constrained Lift-one Algorithm}\label{sec:constrained_lift_one}

For readers' reference, we provide the original lift-one algorithm for general parametric models as Algorithm~3 in Section~S1 of the Supplementary Material. 
As mentioned in the Introduction section, the original lift-one algorithm does not fit our needs for constrained optimal allocation. We provide such an example (Example~\ref{ex:counter}) in Subsection~\ref{sec:counterexample}.

\subsection{An illustrative example for the lift-one algorithm}\label{sec:counterexample}

In this section, we provide an example such that the allocation found by the original lift-one algorithm is not D-optimal under constraints.

\begin{figure}[ht]
\begin{center}
\begin{tikzpicture}
    \begin{axis}[
      xmin=-2.05,xmax=2.05,
      ymin=-1.05,ymax=2.05,
      xticklabels=none,
      yticklabels=none,
      xtick=\empty,ytick=\empty,
      axis line style={draw=none},
    ]
        \addplot [
        domain=-1:1, 
        samples=100, 
        ]
        {0-0.5};  
        \addplot [
        domain=-1:0, 
        samples=100, 
        ]
        {sqrt(3)*x+sqrt(3)-0.5}; 
        \addplot [
        domain=0:1, 
        samples=100, 
        ]
        {-sqrt(3)*x + sqrt(3)-0.5};
        \addplot[samples=50, 
        smooth, 
        domain=-1:1, 
        dotted] coordinates {(0,0-0.5)(0,1.73205-0.5)}; 
        \addplot [name path = A,
        domain = -1:0,
        samples=100,
        dashed
        ]
        {2/3*sqrt(3)*x + 2/3*sqrt(3)-0.5}; 
        \addplot [name path = B,
        domain = -2/3:1/6,
        samples=100,
        dashed
        ]
        {sqrt(3)*x + 2/3*sqrt(3)-0.5}; 
        \addplot [name path = C,
        domain = -7/15:7/15,
        samples=100,
        dashed
        ]
        {7/15*sqrt(3)-0.5}; 
        \addplot [mark=*, mark options={draw=black}, only marks]
        coordinates {
        (-1,0-0.5) 
        };
        \node at (axis cs:-1.3,-0.2-0.5) {$\mathbf{w_2}=(0,1,0)$}; 
        
        \addplot [mark=*, mark options={draw=black}, only marks]
        coordinates {
        (1,0-0.5) 
        };
        \node at (axis cs:1,-0.2-0.5) {$\mathbf{w_1}=(1,0,0)$}; 
        
        \addplot [mark=*, mark options={draw=black}, only marks]
        coordinates {
        (0,1.73205-0.5) 
        };
        \node at (axis cs:0,1.95-0.5) {$\mathbf{w_3}=(0,0,1)$}; 
        
        \addplot [mark=*, mark options={draw=black}, only marks]
        coordinates {
        (0,1.1547-0.5) 
        };
        \node at (axis cs:0.2,1.1547-0.5) {$\mathbf w_a$}; 
        
        \addplot [mark=*, mark options={draw=black}, only marks]
        coordinates {
        (0,0.57735-0.5) 
        };
        \node at (axis cs:0.25,0.5-0.5) {$\mathbf w_o$}; 
        
        \addplot [mark=*, mark options={draw=black}, only marks]
        coordinates {
        (-0.3,0.80829-0.5) 
        };
        \node at (axis cs:-0.4,0.95-0.5) {$\mathbf w_b$};
        
        \addplot [mark=*, mark options={draw=black}, only marks]
        coordinates {
        (-0.2,0.80829-0.5) 
        };
        \node at (axis cs:-0.15,0.7-0.5) {$\mathbf w_d$};

        \addplot [mark=*, mark options={draw=black}, only marks]
        coordinates {
        (0, 0.80829-0.5) 
        };
        \node at (axis cs:0.2,0.92-0.5) {$\mathbf w_c$};
        \addplot [teal!30] fill between [of = A and B, soft clip={domain=-0.2:0}];
        \addplot [teal!30] fill between [of = A and C, soft clip={domain=-0.3:-0.2}];
        
    \end{axis}
   
  \end{tikzpicture}
  \caption{2D Display of Example~\ref{ex:counter}} \label{fig:ex:counter}
  \end{center}
  \end{figure}
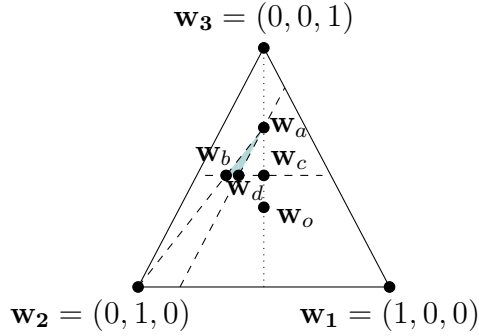

\begin{example}\label{ex:counter}{\rm
Consider an experiment with the logistic regression model $g(\mu_i)=\log(\frac{\mu_i}{1-\mu_i}) = \beta_0 + \beta_1 x_{i1} + \beta_2 x_{i2}$ with $\mu_i = E(Y_i\mid {\mathbf x}_i)$ and ${\mathbf x}_i = (x_{i1}, x_{i2})^T \in \{(-1, -1), (-1, +1), (+1, -1)\}$. In this case, $f({\mathbf w}) = w_1w_2w_3$ up to a constant $C>0$ (see Section~\ref{sec:doptimal} for more details).

When there is no constraint, as a direct conclusion of the inequality of arithmetic and geometric means, $f({\mathbf w})$ attains its global maximum at ${\mathbf w}_o = (\frac{1}{3}, \frac{1}{3}, \frac{1}{3})^T \in S_0$ (see Figure~\ref{fig:ex:counter} for a 2D display).

Suppose we consider a constrained D-optimal design problem with $S = \{(w_1, w_2, w_3)^T \in S_0 \mid w_1 \leq \frac{1}{6}, w_3 \geq \frac{8}{15}, 4w_1 \geq w_3\}$, which is a triangle with vertices ${\mathbf w}_a = (\frac{1}{6}, \frac{1}{6}, \frac{2}{3})^T$, ${\mathbf w}_b = (\frac{2}{15}, \frac{1}{3}, \frac{8}{15})^T$ and ${\mathbf w}_d = (\frac{1}{6}, \frac{3}{10}, \frac{8}{15})^T$ (see the shaded region in Figure~\ref{fig:ex:counter}). 

For illustrative purposes, we let the original lift-one algorithm (Algorithm~3 in the Supplementary Material) start with ${\mathbf w}_a \in S$. With the order $\{1, 3, 2\}$ of $i$, we follow Steps~$3^\circ \sim 5^\circ$ of Algorithm~3 with the ranges of $z$ adjusted according to $S$ (see also Steps~$3^\circ \sim 5^\circ$ of Algorithm~\ref{algo:constrained_general_lift_one} in Subsection~\ref{sec:constrained_algorithm}). At $i=1$, $f_1(z) = \frac{4}{25}z(1-z)^2$ with $z \in \{\frac{1}{6}\}$, which leads to ${\mathbf w}_*^{(1)} = {\mathbf w}_1(\frac{1}{6}) = {\mathbf w}_a$; at $i=3$, $f_3(z) = \frac{1}{4} z (1-z)^2$ with $z \in  \{\frac{2}{3}\}$, which leads to ${\mathbf w}_*^{(3)} = {\mathbf w}_3(\frac{2}{3}) = {\mathbf w}_a$; and at $i=2$, $f_2(z) = \frac{4}{25} z (1-z)^2$ with $z \in [\frac{1}{6}, \frac{1}{3}]$, which is maximized at $z_*=\frac{1}{3}$ and leads to ${\mathbf w}_*^{(2)} = {\mathbf w}_2(z_*) = {\mathbf w}_b$~. That is, after the first round of iterations, ${\mathbf w}_a$ is updated by ${\mathbf w}_b$~. 

We continue the lift-one iterations with ${\mathbf w}_b$~. At ${\mathbf w}_b$, $f_1(z) = \frac{40}{169}z(1-z)^2$ with $z \in \{\frac{2}{15}\}$, which leads to ${\mathbf w}_*^{(1)} = {\mathbf w}_1(\frac{2}{15}) = {\mathbf w}_b$; $f_3(z) = \frac{10}{49} z (1-z)^2$ with $z \in \{\frac{8}{15}\}$, which leads to ${\mathbf w}_*^{(3)} = {\mathbf w}_3(\frac{8}{15}) = {\mathbf w}_b$; and $f_2(z) = \frac{4}{25} z (1-z)^2$ with $z \in [\frac{1}{6}, \frac{1}{3}]$, which is maximized at $z_* = \frac{1}{3}$ and leads to ${\mathbf w}_*^{(2)} = {\mathbf w}_2(\frac{1}{3}) = {\mathbf w}_b$~. That is, the lift-one algorithm converges at ${\mathbf w}_b$~. However, instead of ${\mathbf w}_b$, ${\mathbf w}_d$ is the D-optimal allocation in $S$ (see Example~\ref{ex:counter_continued}).
\hfill{$\Box$}
}\end{example}

\subsection{New algorithm for constrained D-optimal allocations}\label{sec:constrained_algorithm}

To find D-optimal allocations under constraints, we develop a new algorithm, called the {\it constrained lift-one algorithm}, for finding D-optimal allocations in a closed and convex $S$. If $S$ itself is not convex but a finite union $\cup_{k=1}^K S_k$ of closed and convex $S_k$'s, the proposed algorithm can be applied to each $S_k$ and find the D-optimal allocation ${\mathbf w}_k$ in $S_k$~. Then the D-optimal allocation in $S$ is simply the best one among ${\mathbf w}_k$'s.

\begin{algorithm}\label{algo:constrained_general_lift_one}
{\bf Constrained lift-one algorithm under a general setup}
\begin{itemize}
	\item[$1^\circ$] Start with an arbitrary allocation ${\mathbf w}_a=(w_1,\ldots,w_m)^T \in S$ satisfying $f({\mathbf w}_a) > 0$ and $0\leq w_i < 1$, $i=1, \ldots, m$.
	\item[$2^\circ$] Set up a random order of $i$ going through $\{1,2,\ldots,m\}$. For each $i$, do steps $3^\circ\sim 5^\circ$.
	\item[$3^\circ$] For $z \in [0,1]$, let	${\mathbf w}_i(z) = \left(\frac{1-z}{1-w_i} w_1, \ldots, \allowbreak \frac{1-z}{1-w_i} w_{i-1}, z, \frac{1-z}{1-w_i} w_{i+1}, \ldots,\right.$ $\left.\frac{1-z}{1-w_i} w_m\right)^T$ and $f_i(z)$ $=$ $f\{{\mathbf w}_i(z)\}$. Determine $0\leq r_{i1} \leq r_{i2} \leq 1$, such that, ${\mathbf w}_i(z) \in S$ if and only if $z \in [r_{i1}, r_{i2}]$.
	\item[$4^\circ$] Use an analytic solution or the quasi-Newton algorithm to find $z_*$ maximizing $f_i(z)$ with $z \in [r_{i1}, r_{i2}]$. Define ${\mathbf w}_*^{(i)} = {\mathbf w}_i(z_*)$. Note that $f({\mathbf w}_*^{(i)}) = f_i(z_*)$.
	\item[$5^\circ$] If $f({\mathbf w}_*^{(i)}) > f({\mathbf w}_a)$, replace ${\mathbf w}_a$ with ${\mathbf w}_*^{(i)}$, and $f\left({\mathbf w}_a\right)$ with $f({\mathbf w}_*^{(i)})$. 
	\item[$6^\circ$] Repeat Steps $2^\circ\sim 5^\circ$ until convergence, that is, $f({\mathbf w}_*^{(i)}) \leq f({\mathbf w}_a)$ for each $i$. Denote ${\mathbf w}_* = (w_1^*, \ldots, w_m^*)^T$ as the converged allocation. 
	\item[$7^\circ$] Calculate $f_i'(w_i^*)$ for each $i$. If $f_i'(w_i^*) \leq 0$ for all $i$, then go to Step~$10^\circ$. Otherwise, go to Step~$8^\circ$.
	\item[$8^\circ$] Find ${\mathbf w}_o \in {\rm argmax}_{{\mathbf w} \in S} g({\mathbf w})$, where $g({\mathbf w}) = \sum_{i=1}^m w_i (1-w_i^*) f_i'(w_i^*)$ is a linear function of ${\mathbf w} = (w_1, \ldots, w_m)^T$. If $\mathbf w_o$ is not unique, choose any of them. If $g({\mathbf w}_o) \leq 0$, go to Step $10^\circ$. Otherwise, go to Step $9^\circ$.
	\item[$9^\circ$] Use an analytic solution or the quasi-Newton algorithm to find $\alpha_*$ maximizing $h(\alpha) = f\left\{(1-\alpha) {\mathbf w}_* + \alpha {\mathbf w}_o\right\}$ with $\alpha \in [0,1]$ (see Theorem~\ref{thm:maximize_h_alpha}). Let ${\mathbf w}_a =  (1-\alpha_*) {\mathbf w}_* + \alpha_* {\mathbf w}_o$ and go back to Step $2^\circ$.
	\item[$10^\circ$] Report ${\mathbf w}_*$ as the D-optimal allocation.
\end{itemize}
\end{algorithm}

Compared with the original lift-one algorithm, Steps $1^\circ \sim 6^\circ$ in Algorithm~\ref{algo:constrained_general_lift_one} are essentially the same except for the intervals $[r_{i1}, r_{i2}]$ in Step~$3^\circ$ due to constraints. Steps $7^\circ\sim 9^\circ$ in Algorithm~\ref{algo:constrained_general_lift_one} are new. Since the lift-one algorithm utilizes the directional derivatives, the searches for optimal allocations are restricted to the directions between the current allocation and the vertices of $S_0$~. It works for unconstrained optimal design problems but not for constrained ones, since the optimal allocation may not be covered by the directions under constraints (see Example~\ref{ex:counter}). In this case, Steps $7^\circ$ and $8^\circ$ check whether the current allocation is D-optimal. If not, we use Step~$9^\circ$ to adjust the starting allocation and return searching. Theoretical justifications and more details about Steps~$7^\circ \sim 9^\circ$ can be found in Sections~\ref{sec:optimality_general}$\sim$\ref{sec:maximize_h_alpha}.
\hfill{$\Box$}

To find $r_{i1}$ and $r_{i2}$ in Step~$3^\circ$ of Algorithm~\ref{algo:constrained_general_lift_one} in general, we suggest the following procedure: (1) Suppose $0\leq w_i < 1$, we start with the interval $I_0 = [0,1]$ and the line segment $L_0 = \{ {\mathbf w}_i(z) \mid z \in I_0\} \subset S_0$; (2) since $S$ is closed and convex, then $L_0 \cap S$ must be closed and convex as well and remain a line segment in the form of $L_i = \{{\mathbf w}_i(z) \mid z\in I_i\}$ with a closed interval $I_i \subseteq [0,1]$. Then $[r_{i1}, r_{i2}] = I_i$~.  

A special case is that, as all the examples provided in this paper, $S$ can be written in the form of $\{{\mathbf w} \in S_0 \mid {\mathbf a}_\lambda^T {\mathbf w} \leq b_\lambda, \lambda \in \Lambda\}$ with ${\mathbf a}_\lambda \in \mathbb{R}^m$ and $b_\lambda \in \mathbb{R}$. For each $\lambda \in \Lambda$, ${\mathbf a}_\lambda^T {\mathbf w}_i(z) \leq b_\lambda$ with $z\in [0,1]$ implies $\{{\mathbf w}_i(z) \mid z \in I_\lambda\}$ with some interval $I_\lambda \subseteq [0,1]$. Then $[r_{i1}, r_{i2}] =  \cap_{\lambda \in \Lambda} I_\lambda$, which is nonempty (see  Examples~9 and  10 in the Supplementary Material).

\begin{example}\label{ex:counter_continued}
Example~\ref{ex:counter} is considered here again. Recall that ${\mathbf w}_b = (\frac{2}{15}, \frac{1}{3}, $ $\frac{8}{15})^T$ is reported as the converged allocation in Step~$6^\circ$ of Algorithm~3 (or Algorithm~\ref{algo:constrained_general_lift_one}). To check the conditions in Step~$7^\circ$ of Algorithm~\ref{algo:constrained_general_lift_one} for ${\mathbf w}_b$, we obtain 
$f'_1(\frac{2}{15}) = \frac{8}{65} >0$, $f'_2(\frac{1}{3}) = 0$, and $f'_3(\frac{8}{15}) = -\frac{2}{35} <0$.
We then go to Step~$8^\circ$ with $g({\mathbf w}) = \frac{2}{75} (4w_1 - w_3)$. Since $S$ is the convex hull of its vertex set $\{{\mathbf w}_a, {\mathbf w}_b, {\mathbf w}_d\}$, it can be verified that (see Theorem~\ref{thm:max_gw_vextex} in Subsection~\ref{sec:maximize_gw}) ${\mathbf w}_d = (\frac{1}{6}, \frac{3}{10}, \frac{8}{15})^T$ maximizes $g({\mathbf w})$, ${\mathbf w} \in S$. Since $g({\mathbf w}_d) = \frac{4}{1125} > 0$, we go to Step~$9^\circ$ and define $h(\alpha) = f\{(1-\alpha) {\mathbf w}_b + \alpha {\mathbf w}_d\}=\frac{2}{15^3} (4+\alpha) (10 - \alpha)$. Since $h'(\alpha) = \frac{4}{15^3} (3-\alpha) > 0$ for all $\alpha \in [0,1]$, then $\alpha_* = {\rm argmax}_{{\alpha}\in [0,1]} h(\alpha) = 1$. We let ${\mathbf w}_a^{(1)} =  (1-\alpha_*) {\mathbf w}_b + \alpha_* {\mathbf w}_d={\mathbf w}_d$ and go back to Step~$2^\circ$ of Algorithm~\ref{algo:constrained_general_lift_one}. 

First of all, ${\mathbf w}_d$ is a converged allocation in Step~$6^\circ$. Actually, $f_1(z) = \frac{144}{625}z(1-z)^2$ with $z \in [\frac{4}{29},\frac{1}{6}]$ and is maximized at $z_*=\frac{1}{6}$; $f_2(z) = \frac{80}{441}z(1-z)^2$ with $z \in \{\frac{3}{10}\}$ and thus is maximized at $z_*=\frac{3}{10}$; and $f_3(z) = \frac{45}{196}z(1-z)^2$ with $z \in [\frac{8}{15},\frac{10}{17}]$ and is maximized at $z_*=\frac{8}{15}$.  Secondly, since $f_1'(\frac{1}{6}) = \frac{12}{125} > 0$, $f_2'(\frac{3}{10}) = \frac{4}{315} > 0$, and $f_3'(\frac{8}{15}) = -\frac{9}{140} < 0$, we go to Step~$8^\circ$. Since  $g({\mathbf w}) = \frac{2}{25}(w_1 + \frac{w_2}{9} - \frac{3}{8}w_3)$ is maximized at ${\mathbf w}_d$ (see Theorem~\ref{thm:max_gw_vextex} again) and $g({\mathbf w}_d)=0$, we go to Step~$10^\circ$ and report ${\mathbf w}_d$ as the D-optimal allocation.
\hfill{$\Box$}
\end{example}

Given an approximate allocation ${\mathbf w} = (w_1, \ldots, w_m)^T \in S$, if all additional constraints for $S$ take the form of $\sum_{i=1}^m a_i w_i \leq c$ with $a_i\geq 0$ and $c>0$ such as in Examples~\ref{ex:2*3} and \ref{ex:trauma}, we develop the following constrained round-off algorithm to obtain an exact allocation ${\mathbf n} = (n_1, \ldots, n_m)^T$ satisfying ${\mathbf n}/n \in S$ and $\sum_{i=1}^m n_i \leq n$.

\begin{algorithm}\label{algo:constrained_round_off}
{\bf Constrained round-off algorithm for obtaining a feasible exact allocation}
\begin{itemize}
 \item[$1^\circ$] First let $n_i = \lfloor n w_i\rfloor$, the largest integer no more than $n w_i$, $i=1, \ldots, m$, and $k= n - \sum_{i=1}^m n_i$~. Denote $I = \{ i\in \{1, \ldots, m\} \mid w_i > 0, (n_1, \ldots, n_{i-1}, n_i+1, n_{i+1}, \ldots, n_m)/n \in S\}$.
 \item[$2^\circ$] While $k>0$ and $I \neq \emptyset$, do
 \begin{itemize}
 \item[2.1] For $i \in I$, calculate $d_i = f(n_1, \ldots, n_{i-1}, n_i+1, n_{i+1}, \ldots, n_m)$.
 \item[2.2] Pick up any $ i_* \in {\rm argmax}_{i\in I} d_i$~.
 \item[2.3]
 Let $n_{i_*} \leftarrow n_{i_*} + 1$ and $k \leftarrow k-1$. 
 \item[2.4] If $(n_1, \ldots, n_{i_*-1}, n_{i_*}+1, n_{i_*+1}, \ldots, n_m)/n \notin S$, then $I \leftarrow I\setminus \{i_*\}$.
 \end{itemize}
 \item[$3^\circ$] Output ${\mathbf n} = (n_1, \ldots, n_m)^T$.
\end{itemize}
\end{algorithm}

The allocation obtained by Algorithm~\ref{algo:constrained_general_lift_one} with $f({\mathbf w}) = |\sum_{i=1}^m w_i {\mathbf F}_i|$ is known as a {\it locally D-optimal} allocation since it may require assumed values of $\boldsymbol\theta$. With a specified prior distribution $h(\boldsymbol\theta)$ on the parameter space $\boldsymbol\Theta$, 
we may replace $f({\mathbf w})$ with $f_{\rm EW}({\mathbf w}) = |\sum_{i=1}^m w_i E({\mathbf F}_i)|$ and the obtained allocation by Algorithm~\ref{algo:constrained_general_lift_one} is called an {\it EW D-optimal} allocation (see Section~\ref{sec:constrained_optimal_allocation}).

\subsection{D-optimality of Algorithm~\ref{algo:constrained_general_lift_one}}\label{sec:optimality_general}

In this section, we show that the allocation reported by Algorithm~\ref{algo:constrained_general_lift_one} is D-optimal. Throughout this section, we assume that $S$ is closed and convex.

\begin{lemma}\label{lem:f_i_convex}
Suppose $f({\mathbf w}_a) > 0$ for some ${\mathbf w}_a = (w_1, \ldots, w_m)^T \in S$ with $0\leq w_i <1$, $i=1, \ldots, m$. Let $f_i(z) = f\{{\mathbf w}_i(z)\}$ as defined in Algorithm~\ref{algo:constrained_general_lift_one}. Then $\log f_i(z)$ is a concave function on $[r_{i1}, r_{i2}]$. Furthermore, suppose $z_*$ maximizes $f_i(z)$ on $[r_{i1}, r_{i2}]$.   Then (1) if $z_* = r_{i1} < r_{i2}$, then $f_i'(z_*) \leq 0$; (2) if $z_* = r_{i2} > r_{i1}$, then $f_i'(z_*) \geq 0$; and (3) if $z_* \in (r_{i1}, r_{i2})$, then $f_i'(z_*)=0$.
\end{lemma}

\begin{theorem}\label{thm:wi<ri2}
Suppose $f({\mathbf w}) > 0$ for some ${\mathbf w} \in S$. Let ${\mathbf w}_* = (w_1^*, \ldots, w_m^*)^T$ $\in S$ be a converged allocation in Step~$6^\circ$ of Algorithm~\ref{algo:constrained_general_lift_one} with $0\leq w_i^* < 1$, $i=1, \ldots, m$. If $f_i'(w_i^*)\leq 0$ for each $i$, then ${\mathbf w}_*$ must be D-optimal in $S$.
\end{theorem}

Using Lemma~\ref{lem:f_i_convex}, Theorem~\ref{thm:wi<ri2} and the following corollary, the D-optimality of the converged allocation ${\mathbf w}_*$ under some conditions can be easily justified.

\begin{corollary}\label{col:S=S0_fi'<=0}
Suppose ${\rm rank}({\mathbf F}_i) < p$ for each $i$ and $f({\mathbf w}) > 0$ for some ${\mathbf w} \in S$. Let ${\mathbf w}_*\in S$ be a converged allocation in Step~$6^\circ$ of Algorithm~\ref{algo:constrained_general_lift_one}. 
If $w_i^* < r_{i2}$ for each $i$, then ${\mathbf w}_*$ must be D-optimal in $S$.
\end{corollary}

\begin{remark}\label{ex:S=S0_r1_r2}{\rm 
If $S=S_0$, then $[r_{i1}, r_{i2}]$ in Step~$3^\circ$ of Algorithm~\ref{algo:constrained_general_lift_one} is $[0,1]$ for each $i$.
Let ${\mathbf w}_* = (w_1^*, \ldots, w_m^*)^T$ $\in S_0$ be a converged allocation in Step~$6^\circ$ of Algorithm~\ref{algo:constrained_general_lift_one}.
If ${\rm rank}({\mathbf F}_i) < p$ for each $i$ and $f({\mathbf w}) > 0$ for some ${\mathbf w} \in S$, then $w_i^* < r_{i2}=1$ for each $i$. According to Corollary~\ref{col:S=S0_fi'<=0}, ${\mathbf w}_*$ must be D-optimal in $S_0$~.
That is, the original lift-one algorithm (Algorithm~3 in the Supplementary Material) still works for the case $S=S_0$~. 
\hfill{$\Box$}
}\end{remark}

\begin{theorem}\label{thm:constrained_general_optimal}
Suppose ${\rm rank}({\mathbf F}_i) < p$ for each $i$ and  $f({\mathbf w}) > 0$ for some ${\mathbf w} \in S$. Then ${\mathbf w}_*$ reported in Step~$10^\circ$ of  Algorithm~\ref{algo:constrained_general_lift_one} is D-optimal in $S$.
\end{theorem}

\subsection{Maximization of $g({\mathbf w})$ in Step~$8^\circ$ of Algorithm~\ref{algo:constrained_general_lift_one}}\label{sec:maximize_gw}

In this section, we provide the solutions maximizing $g({\mathbf w}) = \sum_{i=1}^m w_i (1-w_i^*) f_i'(w_i^*)$ with ${\mathbf w} = (w_1, \ldots, w_m)^T$ $\in S$, where $S \subseteq S_0$ is closed and convex. By letting $a_i = (1-w_i^*) f_i'(w_i^*)$ and ${\mathbf a} = (a_1, \ldots, a_m)^T$,  $g({\mathbf w}) = {\mathbf a}^T {\mathbf w}$ is a linear function of ${\mathbf w} \in S$.

For many applications including all the examples considered in this paper, $S$ is determined by linear conditions or constraints and the maximization of $g({\mathbf w})$ can be written as:
\begin{equation}\label{eq:linear_program_general}
\begin{array}{cl}
{\rm Max} & {\mathbf a}^T {\mathbf w}\\
{\rm subject\, to} & {\mathbf G} {\mathbf w} \preceq {\mathbf h},\> {\mathbf A} {\mathbf w} = {\mathbf b}
\end{array}
\end{equation}
where ${\mathbf G}\in \mathbb{R}^{r\times m}$, ${\mathbf h}\in \mathbb{R}^r$, ${\mathbf A} \in \mathbb{R}^{s\times m}$, ${\mathbf b} \in \mathbb{R}^s$ are known matrices or vectors, and ``$\preceq$'' is componentwise ``$\leq$''. It is known as a {\it linear program} (LP) problem (see, for example, Section~4.3 in \cite{boyd2004convex}) and can be efficiently solved by using, for example, {\tt R} function {\tt lp} in Package {\tt lpSolve}.

For general cases, $S\subseteq S_0$ is closed and convex. Since $S_0 \subset \mathbb{R}^m$ is bounded, $S$ is bounded and thus compact. According to Theorem~5.6 in \cite{lay2007convex}, $S$ is the convex hull of its profile $E$, the set consisting of all extreme points of $S$. Since $g({\mathbf w})$ is linear on $S$ which depends on $\mathbf w_*$, according to Theorem~5.7 in \cite{lay2007convex}, there exists a ${\mathbf w}_o \in E$ such that 
\[
{\mathbf w}_o \in {\rm argmax}_{{\mathbf w} \in E} g({\mathbf w}) = {\rm argmax}_{{\mathbf w} \in S} g({\mathbf w})
\]
In other words, we only need to search ${\mathbf w}_o$ among the profile $E$ of $S$, which is only a subset of the boundary of $S$. According to the proof of Theorem~\ref{thm:wi<ri2}, if we can find a $\mathbf w_c \in S$ such that $f(\mathbf w_c) > f(\mathbf w_*)$, then $g(\mathbf w_c) > 0$. Since $g(\mathbf w_c)$ is the directional derivative of $f(\mathbf w)$ along $\mathbf w_c - \mathbf w_*$, $g(\mathbf w_c) > 0$ implies that there exists at least one point along the direction with higher $f(\mathbf w)$ value, which is not necessary to be $f(\mathbf w_c)$. Following the proof of Theorem~5.7 in \cite{lay2007convex}, we obtain the following convenient results for special cases:

\begin{theorem}\label{thm:max_gw_vextex}
If there exists a $V = \{{\mathbf w}_1, \ldots, {\mathbf w}_k\} \subset S$, such that, $S$ can be rewritten as $\{b_1 {\mathbf w}_1 + \cdots + b_k {\mathbf w}_k\mid b_1 \geq 0, \ldots, b_k \geq 0, \sum_{i=1}^k b_i = 1\}$, then ${\mathbf w}_o \in {\rm argmax}_{{\mathbf w} \in V} g({\mathbf w})$ maximizes $g({\mathbf w})$ on $S$.
\end{theorem}

In other words, if $S$ is the convex hull of a finite set $V$ (such an $S$ is called a {\it polytope} or {\it convex polytope} in the literature; see, for example, Definition~2.24 in \cite{lay2007convex}), we only need to search ${\mathbf w}_o$ among the finite set $V$. Note that the $V$, known as the {\it vertex set}, may not be unique and may not be the profile of $S$.

Given $a_i = (1-w_i^*) f_i'(w_i^*)$, $i=1, \ldots, m$, we call $r_1, \ldots, r_m$ the {\it ranks} of them, if $\{r_1, \ldots, r_m\} = \{1, \ldots, m\}$ and $a_{r_1} \geq a_{r_2} \geq \cdots \geq a_{r_m}$~. For $S$ taking the form as in Example~\ref{ex:2*3}, we have an analytic solution for ${\mathbf w}_o$:

\begin{theorem}\label{thm:solve_g(w)}
Suppose $S = \{{\mathbf w} \in S_0 \mid w_i \leq c_i, i=1, \ldots, m\}$ with $0< c_i\leq 1$, $i=1, \ldots, m$ and $\sum_{i=1}^m c_i \geq 1$. Then a ${\mathbf w}_o = (w_1^o, \ldots, w_m^o)^T$ maximizing $g({\mathbf w}) = \sum_{i=1}^m a_i w_i$ can be obtained as follows: (i) if $\sum_{i=1}^m c_i = 1$, ${\mathbf w}_o = (c_1, \ldots, c_m)^T$; (ii) if $\sum_{i=1}^m c_i > 1$, then $w_i^o = c_i$ if $i \in \{r_1, \ldots, r_k\}$; $1-\sum_{l=1}^k c_{r_l}$ if $i=r_{k+1}$; and $0$ otherwise, where $r_1, \ldots, r_m$ are the ranks of $a_1, \ldots, a_m$ and $k \in \{1, \ldots, m-1\}$ satisfying $\sum_{l=1}^k c_{r_l} \leq 1 < \sum_{l=1}^{k+1} c_{r_l}$~. 
\end{theorem}

\begin{example}\label{ex:counter_continued2}
Example~\ref{ex:counter} is considered here again. In this case, $S = \{(w_1, w_2,$ $ w_3)^T \in S_0 \mid w_1 \leq \frac{1}{6}, w_3 \geq \frac{8}{15}, 4w_1 \geq w_3\}$ with its vertex set $V = \{{\mathbf w}_a, {\mathbf w}_b, {\mathbf w}_d\}$ (see Figure~\ref{fig:ex:counter}). As mentioned in Example~\ref{ex:counter_continued}, $g({\mathbf w})$ defined with both ${\mathbf w}_b$ and ${\mathbf w}_d$ is maximized in $S$ at ${\mathbf w}_d$, one of the three vertices.
\hfill{$\Box$}
\end{example}

\subsection{Maximization of $h(\alpha)$ in Step~$9^\circ$ of Algorithm~\ref{algo:constrained_general_lift_one}}\label{sec:maximize_h_alpha}

Suppose ${\mathbf w}_* = (w_1^*, \ldots, w_m^*)^T \in S$, $0\leq w_i^* < 1$ for all $i$, is a converged allocation in Step~$6^\circ$ of Algorithm~\ref{algo:constrained_general_lift_one}, and ${\mathbf w}_o = (w_1^o, \ldots, w_m^o)^T \in S$ is obtained in Step~$8^\circ$ with $g({\mathbf w}_o) > 0$. We provide the following results to find $\alpha_*$ maximizing $h(\alpha) = f\left\{(1-\alpha) {\mathbf w}_* + \alpha {\mathbf w}_o\right\}$ in Step~$9^\circ$.

\begin{lemma}\label{lem:h_alpha}
The function $h(\alpha)$ in Step~$9^\circ$ of Algorithm~\ref{algo:constrained_general_lift_one} can be written as
\begin{equation}\label{eq:h_alpha}
h(\alpha) = c_0 + c_1\alpha + \cdots + c_{p-1} \alpha^{p-1} + c_p \alpha^p
\end{equation}
where $c_0 = f({\mathbf w}_*)$, $(c_1, \ldots, c_p)^T = {\mathbf B}_p ^{-1} (h(\frac{1}{p})-c_0, \ldots, h(\frac{p-1}{p}) - c_0, h(1)-c_0)^T$, and ${\mathbf B}_p$ is a $p\times p$ matrix with its $(s,t)$th entry $(\frac{s}{p})^t$.
\end{lemma}

Based on Lemma~\ref{lem:h_alpha}, we can determine the coefficients of $h(\alpha)$ with $h(\frac{1}{p}), \ldots, h(\frac{p-1}{p})$, and $h(1) = f({\mathbf w}_o)$, and then calculate $h'(\alpha)$ by
\begin{equation}\label{eq:h'_alpha}
h'(\alpha) = c_1 + 2 c_2 \alpha + \cdots + (p-1) c_{p-1} \alpha^{p-2} + p c_p \alpha^{p-1}
\end{equation}

\begin{theorem}\label{thm:maximize_h_alpha}
Suppose $f({\mathbf w}_*) > 0$, $g({\mathbf w}_o) = \sum_{i=1}^m w_i^o (1-w_i^*) f_i'(w_i^*) > 0$, $h(\alpha) = f\{(1-\alpha) {\mathbf w}_* + \alpha {\mathbf w}_o\}$,  and $\alpha_*$ maximizes $h(\alpha)$ on $[0,1]$ as defined in Step~$9^\circ$ of Algorithm~\ref{algo:constrained_general_lift_one}. Then {\it (i)} $h(\alpha) > 0$ for all $\alpha \in [0,1)$; {\it (ii)} $h'(0) > 0$; {\it (iii)} if $h(1)>0$ and $h'(1) \geq 0$, then $\alpha_* = 1$; {\it (iv)} if $h(1)>0$ and $h'(1) < 0$, or $h(1)=0$, then there exists a unique $\alpha_0 \in (0,1)$ such that $h'(\alpha_0)=0$, which implies $\alpha_*=\alpha_0$~. By combining (iii) and (iv), we know that $\alpha_*$ always exists and is unique.
\end{theorem}

Based on Theorem~\ref{thm:maximize_h_alpha} and Equation~\eqref{eq:h'_alpha}, if $h'(1) < 0$, one may use, for example, the {\tt R} function {\tt uniroot}, to find $\alpha_* \in (0,1)$ numerically. To avoid numerical errors, we need to double check $\alpha_*\neq 1$ when $f({\mathbf w}_o) < f({\mathbf w}_*)$.

In Step~$9^\circ$ of Algorithm~\ref{algo:constrained_general_lift_one}, we let ${\mathbf w}_a = (1-\alpha_*) {\mathbf w}_* + \alpha_* {\mathbf w}_o$~. According to Theorem~\ref{thm:maximize_h_alpha} (ii), $h'(0) > 0$, which implies $f({\mathbf w}_a) = h(\alpha_*) > f(\mathbf w_*)$. Nevertheless, it does not mean that ${\mathbf w}_a$ is D-optimal already.

\section{D-optimal Samplers for Generalized Linear Models}
\label{sec:doptimal}

In this section, we utilize local and EW D-optimal samplers for univariate responses, such as in Example~\ref{ex:2*3}. Recall that we assign $n_i$ participants to the $i$th category. We let $Y_{ij}$ stand for the univariate response of the $j$th participant of the $i$th category. Generalized linear models \citep{pmcc1989, dobson2018}
\begin{equation}\label{eq:glm}
E(Y_{ij}\mid {\mathbf x}_i) = \mu_i \mbox{ and } g(\mu_i) = \eta_i = {\mathbf h}({\mathbf x}_i)^T \boldsymbol\theta
\end{equation}
have been widely used, where $i=1, \ldots, m$; $j=1, \ldots, n_i$; $g$ is a given ({\it link}) function; $\eta_i$ is known as a linear predictor; 
${\mathbf h}({\mathbf x}_i) = (h_1({\mathbf x}_i), \ldots, h_p({\mathbf x}_i))^T$ are $p$ given predictor functions; and $\boldsymbol\theta=(\theta_1, \ldots, \theta_p)^T$ are the regression coefficients. Commonly used generalized linear models (GLM, see Table~5 in the Supplementary Material) cover Gaussian response (that is, linear models), binary response (Bernoulli, such as Example~\ref{ex:2*3}), count response (Poisson), and real positive response (Gamma, Inverse Gaussian).

Assuming that $Y_{ij}$'s are independent, the Fisher information matrix (see, for example, \cite{ym2015}) 
\[
{\mathbf F}({\mathbf w}) = n \sum_{i=1}^m w_i {\mathbf F}_i = n \sum_{i=1}^m w_i \nu_i {\mathbf h}({\mathbf x}_i) {\mathbf h}({\mathbf x}_i)^T = n{\mathbf X}^T {\mathbf W} {\mathbf X}
\] 
where ${\mathbf w} = (w_1, \ldots, w_m)^T$, $w_i = n_i/n$, ${\mathbf X} = ({\mathbf h}({\mathbf x}_1), \ldots, {\mathbf h}({\mathbf x}_m))^T$ is an $m\times p$ matrix, ${\mathbf W} = {\rm diag} \{w_1 \nu_1, \ldots, w_m \nu_m\}$, and $\nu_i = \nu(\eta_i) = (\partial \mu_i/\partial\eta_i)^2/{\rm Var}(Y_{ij})$, $i=1, \ldots, m$. We provide examples of $\nu(\eta_i)$ for commonly used GLMs in Table~5 of the Supplementary Material. For GLMs, $f({\mathbf w}) = |{\mathbf X}^T {\mathbf W} {\mathbf X}|$ and $f_{\rm EW}({\mathbf w}) = |{\mathbf X}^T E({\mathbf W}) {\mathbf X}|$ (see Section~\ref{sec:constrained_optimal_allocation}).

According to Lemma~4.1 in \cite{ym2015}, $f_i(z) = az(1-z)^{p-1} + b(1-z)^p$, where $b=f_i(0)$, $a=\{f({\mathbf w})-b(1-w_i)^p\}/\{w_i (1-w_i)^{p-1}\}$ if $w_i > 0$; and $b=f({\mathbf w})$, $a=f_i(1/2) 2^p - b$ otherwise. In both cases, $a\geq 0$, $b\geq 0$, and $a+b>0$. To implement Step~$7^\circ$ of Algorithm~\ref{algo:constrained_general_lift_one}, we need 
\begin{equation}\label{eq:glm_f_i'(x)}
f_i'(z) = \{a - bp + (b-a)pz\} (1-z)^{p-2}
\end{equation}
Here $m\geq p\geq 2$. Similar to Lemma~4.2 in \cite{ym2015}, we provide the following lemma for maximizing $f_i(z)$ with constraints:

\begin{lemma}\label{lem:restrictedliftone}
Denote $l(x)=ax(1-x)^{p-1}+b(1-x)^{p}$ with $a\geq 0$, $b\geq 0$ and $a+b >0$. Let $\delta = (a-bp)/\{(a-b)p\}$ when $a\neq b$. Then
\begin{equation}\label{eq:x_*}
x_* = 
\left\{\begin{array}{cl}
\delta, &\mbox{ if }a > bp\mbox{ and }r_1\leq \delta \leq r_2;\\
r_2,    &\mbox{ if }a > bp\mbox{ and }\delta > r_2;\\
r_1,    &\mbox{ otherwise }.
\end{array}\right.
\end{equation}
maximizes $l(x)$ with constraints $0\leq r_1 \leq x\leq r_2\leq 1$.
\end{lemma}

\begin{example}\label{ex:2*3_continued}
Example~\ref{ex:2*3} is considered here again. In this case, $N=500$ eligible volunteers are available for $m=6$ categories with frequencies $(N_1, N_2, \ldots, N_6) = (50, 40, 10, $ $200, 150, 50)$. For illustration purposes, we consider a logistic regression model (GLM with Bernoulli($\mu_i$) and logit link):
\begin{equation}\label{eq:2*3_main}
{\rm logit} \{P(Y_{ij}=1 \mid x_{i1}, x_{i2})\} = \beta_0 + \beta_1 x_{i1} + \beta_{21} 1_{\{x_{i2}=1\}} + \beta_{22} 1_{\{x_{i2}=2\}} 
\end{equation}
where $i=1, \ldots, 6$; $j=1, \ldots, n_i$; and ${\rm logit}(\mu) = \log\{\mu/(1-\mu)\}$. Model~\eqref{eq:2*3_main} is a main-effects model with gender and age group as factors. 

To sample $n=200$ from $m=6$ categories or strata, the proportionally stratified allocation is ${\mathbf w}_{\rm p} = (0.10, 0.08, 0.02, 0.40, 0.30, 0.10)^T$ or ${\mathbf n}_{\rm p} = (20, 16, 4, 80, 60, 20)^T$, 
while the (constrained) uniformly stratified allocation is ${\mathbf w}_{\rm u} = (0.19, 0.19, 0.05, 0.19, 0.19, 0.19)^T$ or ${\mathbf n}_{\rm u} = (38, 38, 10, 38,$ $38, 38)^T$.
By implementing Algorithms~\ref{algo:constrained_general_lift_one} and ~\ref{algo:constrained_round_off} in {\tt R} with assumed $\boldsymbol\beta = (\beta_0, \beta_1, \beta_{21}, \beta_{22} )^T  = (0,3,3,3)^T$, we obtain the (locally) D-optimal allocation ${\mathbf w}_{\rm o} = (0.25, 0.20,$ $0.05, 0.50,$ $0, 0)^T$ or ${\mathbf n}_{\rm o} = (50, 40, 10, 100, 0, 0)^T$.
Compared with ${\mathbf w}_{\rm o}$, the relative efficiency of ${\mathbf w}_{\rm p}$ is $\{|{\mathbf F} ({\mathbf w}_{\rm p})|/|{\mathbf F}({\mathbf w}_{\rm o})|\}^{1/p} = 53.93\%$ with the number of parameters $p=4$, and the relative efficiency of ${\mathbf w}_{\rm u}$ is $\{|{\mathbf F} ({\mathbf w}_{\rm u})|/|{\mathbf F}({\mathbf w}_{\rm o})|\}^{1/p} = 78.99\%$. Both are much less efficient than ${\mathbf w}_o$~. 

We also look into the robustness of our optimal allocations to model misspecification. Assuming that the true link in this study is not the assumed logit link but probit,  complementary log-log, or log-log link (see Table~5 in the Supplementary Material), we check the relative efficiency of the allocation obtained under logit to the D-optimal allocations based on the true link. Notice that the log-log link shares the same ${\mathbf W}$ matrix as the complementary log-log link. Since $\boldsymbol{\beta}=(0,3,3,3)$ satisfies the conditions of Theorem~3.2 in \cite{ym2015}, the D-optimal designs are saturated and different links lead to the same D-optimal allocation. Therefore, the relative efficiency of our proposed allocation under logit remains $100\%$  with respect to link misspecifications. In Section~S4 of the Supplementary Material, we provide an example with different assumed parameter values, which still have $99\%$ relative efficiencies with link misspecifications. We also provide the results based on the root mean squared errors (RMSE) in Table~6 in the Supplementary Material, which is consistent with the relative efficiency result and confirms the robustness of our proposed allocations.

To compare the accuracy of the estimated regression coefficients based on different samplers, we use the RMSE ($\{\sum_{i\in I} (\hat{\beta}_i - \beta_i)^2/|I|\}^{1/2}$ given an index set $I$). For illustration purposes, we assume that the true parameters are $(\beta_0, \beta_1, \beta_{21}, \beta_{22} ) = (0, 3, 3, 3)$ and run $100$ simulations. In each simulation, we generate $N=500$ independent observations based on Model~\eqref{eq:2*3_main} and use SRSWOR, proportionally stratified sampler, uniformly stratified sampler, and D-optimal sampler, respectively, to sample $n=200$ observations out of $500$. We then fit Model~\eqref{eq:2*3_main} using the $n=200$ observations to get the estimated parameters $(\hat{\beta}_0, \hat{\beta}_1, \hat{\beta}_{21}, \hat{\beta}_{22})$.
The average and standard deviation (sd) of RMSEs across 100 simulations are listed in Table~\ref{tab:example2.1_model2}. According to the RMSE with index set $I = \{1, 21, 22\}$ (column ``all except $\beta_0$'' in Table~\ref{tab:example2.1_model2}), SRSWOR is the least accurate, proportional stratified sampler is a little better, uniformly stratified sampler is much better, and locally D-optimal sampler is the best, which is much closer to the full data estimates. For readers' reference, we also list the RMSEs for individual $\beta_i$'s.

If we know something about $\boldsymbol\theta$ but not their exact values, we recommend EW D-optimal samplers instead (see also Section~\ref{sec:constrained_optimal_allocation}). For illustration purposes, we consider three different prior distributions: (i) uniform prior: $\boldsymbol\theta \sim {\rm Unif}(-2,2) \times {\rm Unif}(-1,5) \times {\rm Unif}(-1,5) \times {\rm Unif}(-1,5)$; (ii) normal prior: $h(\boldsymbol\theta) = \phi\left(\frac{\beta_0}{0.5}\right) \times 
 \phi\left(\frac{\beta_1-2}{0.5}\right) \times \phi\left(\frac{\beta_{21}-2}{0.5}\right) \times \phi\left(\frac{\beta_{22}-2}{0.5}\right)$; (iii) Gamma prior: $\boldsymbol\theta \sim N(0,1) \times {\rm Gamma}(1, 2)$ $ \times {\rm Gamma} (1, 2) \times {\rm Gamma}(1, 2)$. 
The relevant expectations $E\left[\nu\{{\mathbf h}^T({\mathbf x}_i) \boldsymbol\theta\}\right]$ can be numerically computed using, for example, {\tt R} function {\tt hcubature} in package {\tt cubature}.
Compared with the locally D-optimal allocation ${\mathbf w}_{\rm o}$ which samples only from four categories, EW allocations are not so extreme,   
such as ${\mathbf w}_{\rm uEW} = (0.240, 0.200, 0.050, 0.211, 0.101,$ $ 0.198)^T$ with the uniform prior, 
${\mathbf w}_{\rm nEW} = (0.250, 0.200, 0.050, 0.334, 0, 0.166)^T$ with the normal prior,
and ${\mathbf w}_{\rm gEW} = (0.240, 0.200, 0.050, 0.214,$ $0.096, 0.200)^T$ with the gamma prior.
Compared with ${\mathbf w}_{\rm o}$, their relative efficiencies are $85.90\%$, $94.96\%$, and $86.32\%$, respectively, which are still much better than SRSWOR, proportionally stratified, and uniformly stratified samplers. In terms of RMSE (see Table~\ref{tab:example2.1_model2}), the conclusions are consistent.
\hfill{$\Box$}
\end{example}

\begin{table}[ht] \caption{Average (sd) of RMSE over $100$ Simulations under Model~\eqref{eq:2*3_main}}\label{tab:example2.1_model2}
\begin{adjustbox}{width={\textwidth},totalheight={\textheight},keepaspectratio}%
\begin{tabular}{ |c|c|c|c|c|c| }
 \hline
 \multirow{2}{4em}{Sampler}&\multicolumn{5}{c|}{Average (sd) of RMSE} \\\cline{2-6}
 & $\beta_0$ & all except $\beta_0$ & $\beta_1$ & $\beta_{21}$ & $\beta_{22}$\\
 \hline
 Full Data & 0.195(0.145) & 6.317(4.070) & 0.363(0.289)  & 2.751(5.468) & 9.098(7.018)\\
 SRSWOR & 0.314(0.216) & 9.984(3.226) & 0.917(2.543) & 8.098(7.885) &  12.976(5.245)\\
 Proportionally Stratified & 0.412(0.304) & 9.496(3.682) & 1.016(2.545) & 7.311(7.942) & 12.469(5.682) \\
 Uniformly Stratified & 0.235(0.193)  &  7.967(4.659) & 3.855(6.673) & 3.353(6.254) & 9.657(7.297)  \\
 Locally D-opt & 0.202(0.150) & 7.103(4.098) & 0.485(0.438)  &  3.890(6.507) & 9.883(6.821)  \\
  Unif EW D-opt & 0.201(0.145)  & 7.556(4.653) & 1.538(3.942) &  3.920(6.561) &  9.273(7.151)\\
 Normal EW D-opt & 0.202(0.147) & 7.252(4.407) &  1.347(3.664) &  3.982(6.687) &  9.302(7.150) \\
 Gamma EW D-opt & 0.205(0.153) & 7.718(4.476) & 1.535(4.008) & 3.955(6.652)  &  9.585(7.080) \\
 \hline
\end{tabular}
\end{adjustbox}
\end{table}

Known as the {\it uniform} allocation, ${\mathbf w} = (1/m, \ldots, 1/m)^T$ has a special role in optimal design theory, which is recommended for linear models or as a robust design (see, for example, \cite{yang2012optimal}). In this paper, we introduce constrained uniform allocations such as ${\mathbf w}_{\rm u}$ in Example~\ref{ex:2*3}. They are D-optimal for saturated cases (that is, $m=p$).

\begin{lemma}\label{lem:constrained_uniform}
Suppose ${\mathbf w}_* = (w_1^*, \ldots, w_m^*)^T$ maximizes  $f({\mathbf w}) = \prod_{i=1}^m w_i$ under the constraints $0\leq w_i \leq c_i$, $i=1, \ldots, m$ and $\sum_{i=1}^m w_i = 1$, where $0< c_i\leq 1$, $i=1, \ldots, m$ and $\sum_{i=1}^m c_i \geq 1$. Then
(i) if $\min_{1\leq i\leq m} c_i \geq 1/m$, then ${\mathbf w}_* = (1/m, \ldots, 1/m)^T$;
(ii) if $0 \leq \min_{1\leq i\leq m} c_i < 1/m$ and $\sum_{i=1}^m c_i > 1$, then there exist $1\leq k\leq m-1$ and $c_{(k)} \leq u < c_{(k+1)}$, such that, $w_i^* = c_i$ if $c_i\leq u$, and $w_i^* = u$ if $c_i > u$, where $0< c_{(1)} \leq c_{(2)}\leq \cdots \leq  c_{(m)}\leq 1$ are order statistics of $c_1, \ldots, c_m$;
(iii) if $\sum_{i=1}^m c_i = 1$, then $w_i^*=c_i$, $i=1, \ldots, m$.
\end{lemma}

We call the ${\mathbf w}_*$ described in Lemma~\ref{lem:constrained_uniform} a {\it constrained uniform allocation} and the corresponding sampler a (constrained) {\it uniformly stratified} sampler.

\begin{theorem}\label{thm:constrained_uniform}
For GLM~\eqref{eq:glm} with $m=p$, if $S = \{{\mathbf w} \in S_0 \mid w_i \leq c_i, i=1, \ldots, m\}$ with $0< c_i\leq 1$, $i=1, \ldots, m$ and $\sum_{i=1}^m c_i \geq 1$, then the constrained uniform allocation described in Lemma~\ref{lem:constrained_uniform} is both D-optimal and EW D-optimal.
\end{theorem}

\begin{example}\label{ex:modified_example_2.1}{\rm 
Example~\ref{ex:2*3} continues here. If we consider another logistic regression model
\begin{eqnarray}
{\rm logit} \{P(Y_{ij}=1 \mid x_{i1}, x_{i2})\} &=& \beta_0 + \beta_1 x_{i1} + \beta_{21} 1_{\{x_{i2}=1\}} + \beta_{22} 1_{\{x_{i2}=2\}} \nonumber \\
&+ &\  \beta_{121} x_{i1} 1_{\{x_{i2}=1\}} + \beta_{122} x_{i1}1_{\{x_{i2}=2\}}\label{eq:2*3_interaction} 
\end{eqnarray}
for Example~\ref{ex:2*3}, which adds two order-$2$ interactions to Model~\eqref{eq:2*3_main}. Then $m=p=6$. According to Theorem~\ref{thm:constrained_uniform}, the constrained uniform allocation ${\mathbf w}_{\rm u} = (0.19, 0.19, 0.05, 0.19, 0.19,$ $0.19)^T$ is both D-optimal and EW D-optimal. In this case, the uniformly stratified sampler is the same as the D-optimal and EW D-optimal samplers. 

To compare the SRSWOR, proportionally stratified, uniformly stratified/locally D-optimal/EW D-optimal samplers, for illustration purposes, we assume that the true parameters are $(\beta_0, \beta_1, \beta_{21}, \beta_{22}, \beta_{121}, \beta_{122}) = (0, $ $-0.1,$ $-0.5, -2, -0.5, -1)$. We run $100$ simulations using Model~\eqref{eq:2*3_interaction}. In this scenario, 
the proportionally stratified allocation ${\mathbf w}_{\rm p}$ and the uniformly stratified allocation ${\mathbf w}_{\rm u}$ are the same as in Example~\ref{ex:2*3}, and the D-optimal allocation ${\mathbf w}_{\rm o} = {\mathbf w}_{\rm u}$~.
The relative efficiencies of ${\mathbf w}_{\rm p}$ and ${\mathbf w}_{\rm u}$ compared with ${\mathbf w}_{\rm o}$ are $73.30\%$ and $100\%$, respectively. In terms of robustness to model misspecifications, the relative efficiencies with true links as the probit, log-log, and complementary log-log are again $100\%$ due to Theorem~\ref{thm:constrained_uniform}. The average and standard deviation of the $100$ RMSEs are reported in Table~\ref{tab:example2.1_model3}. Again, the D-optimal sampler (same as the uniformly stratified sampler in this scenario) significantly reduces the RMSEs based on SRSWOR or the proportionally stratified sampler. 
}\hfill{$\Box$}
\end{example}

\begin{table}[ht] \caption{Average (sd) of RMSE over 100 Simulations under Model~\eqref{eq:2*3_interaction}}\label{tab:example2.1_model3}
\begin{adjustbox}{width={\textwidth},totalheight={\textheight},keepaspectratio}%
\begin{tabular}{ |c|c|c|c|c|c|c|c|}
 \hline
 \multirow{2}{4em}{Sampler}&\multicolumn{7}{c|}{Average (sd) of RMSEs} \\\cline{2-8}
 & $\beta_0$ & all except $\beta_0$ & $\beta_1$ & $\beta_{21}$ & $\beta_{22}$ & $\beta_{121}$ & $\beta_{122}$\\
 \hline
 Full Data & 0.240(0.206)& 3.753(3.752) & 0.285(0.230) & 0.350(0.263) & 4.622(6.406) & 0.403(0.319) & 5.004(6.343)\\
 SRSWOR & 0.340(0.268) & 6.230(3.348) & 0.386(0.290)& 0.552(0.398) & 9.005(6.826) &0.613(0.508) & 7.184(6.845)\\
 Proportionally Stratified & 0.379(0.316) & 6.347(3.267) & 0.459(0.369) & 0.593(0.512) & 9.400(6.877) & 0.742(0.568) & 6.757(6.904)\\
 Uniformly/D-opt/EW D-opt & 0.267(0.203)& 4.186(3.944) & 0.425(0.304) & 0.367(0.272) & 4.973(6.936) & 0.602(0.497) & 5.485(6.887)\\
 \hline
\end{tabular}
\end{adjustbox}
\end{table}

\section{D-optimal Samplers for Multinomial Logit Models}\label{sec:categorical}

In this section, we utilize D-optimal samplers for categorical responses as in Example~\ref{ex:trauma}. For the $i$th experimental setting ${\mathbf x}_i = (x_{i1}, \ldots, x_{id})^T$, $n_i$ categorical responses are collected i.i.d.~from a discrete distribution with $J\geq 2$ categories, $i=1, \ldots, m$. The summary statistics 
${\mathbf Y}_i=(Y_{i1},\cdots,Y_{iJ})^T \sim {\rm Multinomial}(n_i; \pi_{i1}, $ $\cdots,$ $ \pi_{iJ})$,
where $Y_{ij}$ is the number of responses of the $j$th category, $\pi_{ij}$ is the probability that the response falls into the $j$th category at ${\mathbf x}_i$~. Assuming $\pi_{ij}>0$ for all $i=1, \ldots, m$ and $j=1, \ldots, J$, multinomial logit models have been widely used in the literature (see \cite{bu2020} and references therein), including commonly used baseline-category, cumulative, adjacent-categories, and continuation-ratio logit models.

\begin{example}\label{ex:trauma_continued}
Example~\ref{ex:trauma} is considered here again. In this case, the response has $J=5$ categories, and there are $m=8$ distinct experimental settings determined by $d=2$ factors, {\tt severity} ($x_{i1} \in \{0, 1\}$) and {\tt dose} ($x_{i2} \in \{1,2,3,4\}$). For illustration purposes, we first fit the original data using the four different multinomial logit models with main effects, each with proportional odds ({\it po}) or nonproportional odds ({\it npo}) assumptions \citep{bu2020}. According to the Akaike information criterion (AIC, \cite{AIC1973information}), we choose the cumulative logit model with npo: 
\[
\log\left(\frac{\pi_{i1} + \cdots + \pi_{ij}}{\pi_{i,j+1} + \cdots + \pi_{i5}}\right) = \beta_{j1} + \beta_{j2} x_{i1} + \beta_{j3} x_{i2} 
\]
with $i=1, \ldots, 8$ and $j=1, 2, 3, 4$. 

The fitted parameters $\hat{\boldsymbol\theta} = (\hat\beta_{11}, \hat\beta_{21}, \hat\beta_{31},$ $ \hat\beta_{41}, \hat\beta_{12}, \hat\beta_{22}, \hat\beta_{32}, \hat\beta_{42}, \hat\beta_{13}, \hat\beta_{23},$ $\hat\beta_{33}, \hat\beta_{43})^T$ $=$ $(-4.047, -2.225, -0.302, 1.386, 4.214, 3.519, 2.420,$ $1.284,$ $ -0.131,$ $-0.376,$ $-0.237,$ $-0.120)^T$ is used for finding the locally D-optimal allocation ${\mathbf w}_{\rm o}$ and ${\mathbf n}_{\rm o}$ for choosing $n=600$ participants. Since we do not have true parameter values for real data, in order to design EW D-optimal sampler, following Example~5.2 in \cite{bu2020}, we extract $B=1,000$ bootstrapped samples from the original data and fit the cumulative npo model with bootstrapped samples to obtain randomized parameter vectors $\hat{\boldsymbol\theta}_{(1)}, \ldots, \hat{\boldsymbol\theta}_{(B)}$ serving as an empirical distribution of $\boldsymbol\theta$. Among the fitted parameters by SAS PROC LOGISTIC command, $956$ parameter vectors are feasible, that is, in the parameter space $\boldsymbol\Theta = \{\boldsymbol\theta \in \mathbb{R}^{12} \mid \beta_{j1} + \beta_{j2} x_{i1} + \beta_{j3} x_{i2} < \beta_{j+1,1} + \beta_{j+1,2} x_{i1} + \beta_{j+1,3} x_{i2}, j=1, 2, 3; i=1, \ldots, 8\}$ of cumulative logit model (see Section~5.1 in \cite{bu2020}). We denote them by $\boldsymbol\theta_i$, $i = 1, \ldots, 956$.
Then we replace ${\mathbf F}_i$ with $\hat{E}(\mathbf{F}_i)=\sum_{i=1}^{956}\mathbf{F}_i(\boldsymbol\theta_i)/956$ to obtain the EW D-optimal allocation ${\mathbf w}_{\rm EW}$ and ${\mathbf n}_{\rm EW}$, which maximizes $\mid \sum_{i=1}^{8} w_i \hat{E}(\mathbf{F}_i)\mid$. The corresponding allocations are listed in Table~\ref{tab:trauma_original_allocation}. In Table~\ref{tab:eff_trauma}, we list the quantiles of relative efficiencies of SRSWOR (realized allocations after sampling), proportionally stratified, uniformly stratified, and EW D-optimal allocations with respect to the locally D-optimal allocations based on $\boldsymbol\theta_1, \ldots, \boldsymbol\theta_{956}$, respectively. From Table~\ref{tab:eff_trauma}, we conclude that in this case, the EW D-optimal sampler is highly efficient compared with the locally D-optimal sampler, and both of them are much more efficient than SRSWOR, proportionally or uniformly stratified sampler.
\hfill{$\Box$}
\end{example}

According to Table~\ref{tab:eff_trauma}, the two additional constraints in Example~\ref{ex:trauma}, $n(w_1 + w_2 + w_3 + w_4) \leq 392$ and $n(w_5 + w_6 + w_7 + w_8) \leq 410$, are not attained for locally D-optimal and EW D-optimal allocations. In other words, the constrained D-optimal allocations in Example~\ref{ex:trauma_continued} are the same as unconstrained ones in this case. In Example~12 of the Supplementary Material, we provide another example of the sampling problems with the trauma clinical study where the constraints make a difference.

\begin{table}[ht] \caption{Allocations (Proportions) for Stratified Samplers in Example~\ref{ex:trauma_continued}}\label{tab:trauma_original_allocation}
\tiny
\begin{center}
\begin{tabular}{|c|c|c|c|c|c|c|c|c| }
 \hline
 Severity &\multicolumn{4}{c|}{Mild}&\multicolumn{4}{c|}{Severe} \\\hline
 Dose & 1 & 2 & 3 & 4 & 1 & 2 & 3 & 4\\\hline
 Proportional &\thead{78\\(0.130)} & \thead{70\\(0.117)} & \thead{75\\(0.125)} & \thead{72\\(0.120)} & \thead{79\\(0.132)} & \thead{72\\(0.120)} & \thead{80\\(0.133)}& \thead{74\\(0.123)} \\ \hline
  Uniform &  \thead{75\\(0.125)}& \thead{75\\(0.125)} & \thead{75\\(0.125)} & \thead{75\\(0.125)} & \thead{75\\(0.125)} & \thead{75\\(0.125)} & \thead{75\\(0.125)} & \thead{75\\(0.125)}\\ \hline
  Locally D-opt ($\hat{\boldsymbol\theta}$) & \thead{155\\(0.258)} & \thead{0\\(0)} & \thead{0\\(0)} & \thead{100\\(0.167)} & \thead{168\\(0.280)} & \thead{0\\(0)}& \thead{0\\(0)} & \thead{177\\(0.295)}\\ \hline
  EW D-opt & \thead{147\\(0.245)} & \thead{0\\(0)} & \thead{0\\(0)} & \thead{109\\(0.182)} & \thead{168\\(0.280)} & \thead{0\\(0)} & \thead{0\\(0)}  & \thead{176\\(0.293)} \\ 
 \hline
\end{tabular}
\end{center}
\normalsize
\end{table}

\begin{table}[ht]
\caption{Quantiles of Relative Efficiencies in Example~\ref{ex:trauma_continued}}\label{tab:eff_trauma}
\begin{center}
\begin{tabular}{ |c | c c c c c|}\hline
Sampler & Minimum & 1st Quartile & Median & 3rd Quartile & Maximum \\ \hline
SRSWOR & 76.23\% & 80.16\% & 80.65\% & 81.15\% & 84.11\% \\
Proportional & 77.32\% & 80.33\% & 80.66\% & 80.97\% & 83.39\% \\
Uniform & 77.23\% & 80.05\% & 80.40\% & 80.71\% & 83.13\% \\
EW D-opt & 98.91\% & 99.80\% & 99.90\% & 99.96\% & 100\% \\\hline
\end{tabular}
\end{center}
\end{table}

\section{Conclusion}
\label{sec:conc}

In this paper, we consider the constrained subsampling problem for paid research studies or clinical trials to estimate the treatment effects or regression coefficients as accurately as possible. Typically we have some covariates, such as gender and age, collected along with candidates, which are known to have some influences on the treatment effects. If we do not have any idea about the regression coefficients associated with the covariates, we recommend (constrained) uniformly stratified sampler (see Lemma~\ref{lem:constrained_uniform}); if we have some information about the regression coefficients, such as their signs or ranges, we recommend EW D-optimal sampler; if we have a good idea about the regression coefficients such as estimates from a pilot study, we recommend (locally) D-optimal sampler. We use two examples, one with binary responses and generalized linear models, and the other with 5-category responses and multinomial logistic models, to show that our recommended samplers can be much more efficient than classical samplers for paid research studies or clinical trials. The recommended samplers are fairly robust under model misspecification. We also show that under some circumstances, the constrained uniform sampler is optimal from an experimental design's perspective and can be used as a robust sampling strategy in paid research studies if little information is known about the effects of the covariates.

To find an EW D-optimal sampler or locally D-optimal sampler under constraints, we propose a new algorithm called the constrained lift-one algorithm. While keeping the high efficiency of lift-one algorithms, the proposed algorithm corrects the original lift-one algorithm when additional constraints are added to the design space $S$. Compared with the constrained optimal designs in the literature, our algorithm can provide highly efficient results for more general statistical models under more general constraints including but not limited to $n_i \le N_i$~.


\section*{Supplementary Material}
	
{\bf S1 General lift-one algorithm (without constraints):} The lift-one algorithm for general parametric models without constraints; 
{\bf S2 Commonly used GLM models:} A table that lists commonly used GLM models, the corresponding link functions, and $\nu$ functions;
{\bf S3 Two examples of finding $r_{i1}$ and $r_{i2}$ in Algorithm~\ref{algo:constrained_general_lift_one}:} Two examples with details in finding $r_{i1}$ and $r_{i2}$;
{\bf S4 Another example of robustness under GLM models:} An example assuming different parameter values from Example~\ref{ex:2*3_continued}'s for robustness with respect to model misspecification;
{\bf S5 Another example of trauma clinical study:} An example that the D-optimal allocations attain one of the constraints;
{\bf S6 Proofs:} Proofs for lemmas and theorems in this paper. 

	\par
\section*{Acknowledgments}
	Supported in part by the U.S.~NSF grant DMS-1924859.
	\par
	


\clearpage
\setcounter{page}{1}
\def\thepage{S\arabic{page}}

\fontsize{10.95}{14pt plus.8pt minus .6pt}\selectfont
\vspace{0.8pc}
\centerline{\large\bf Constrained D-optimal Design for Paid Research Study}
\vspace{.25cm}
\centerline{Yifei Huang$^1$, Liping Tong$^2$, Jie Yang$^1$} 
\vspace{.4cm}
\centerline{\it $^1$University of Illinois at Chicago, $^2$Advocate Aurora Health}
\vspace{.55cm}
 \centerline{\bf Supplementary Material}
\vspace{.55cm}
\fontsize{9}{11.5pt plus.8pt minus .6pt}\selectfont
\noindent
{\bf S1 General lift-one algorithm (without constraints):} The lift-one algorithm for general parametric models without constraints; \\
{\bf S2 Commonly used GLM models:} A table that lists commonly used GLM models, the corresponding link functions, and $\nu$ functions;\\
{\bf S3 Two examples of finding $r_{i1}$ and $r_{i2}$ in Algorithm~1:} Two examples with details in finding $r_{i1}$ and $r_{i2}$;\\
{\bf S4 Another example of robustness under GLM models:} An example assuming different parameter values from Example~6's for robustness with respect to model misspecification;\\
{\bf S5 Another example of trauma clinical study:} An example that the D-optimal allocations attain one of the constraints;\\
{\bf S6 Proofs:} Proofs for lemmas and theorems in this paper. 

\par

\setcounter{section}{0}
\setcounter{equation}{0}
\def\thesection{S\arabic{section}}
\numberwithin{equation}{section}

\setcounter{algorithm}{2}
\setcounter{table}{4}
\setcounter{example}{8}
\setcounter{lemma}{5}

\fontsize{12}{14pt plus.8pt minus .6pt}\selectfont

\section{General lift-one algorithm (without constraints)}\label{sec:original_lift_one_algorithm}

For readers' reference, in this section, we provide the lift-one algorithm for general parametric models. The lift-one algorithms for specific models can be found in \cite{ymm2016} for GLMs with binary responses, \cite{ym2015} for general GLMs, \cite{ytm2016} for cumulative link models, and \cite{bu2020} for multinomial logistic models.

\begin{algorithm}\label{algo:original_general_lift_one}
{\bf Original lift-one algorithm under a general setup}
\begin{itemize}
	\item[$1^\circ$] Start with an arbitrary allocation ${\mathbf w}_a=(w_1,\ldots,w_m)^T \in S_0$ satisfying $f({\mathbf w}_a) > 0$ and $0\leq w_i < 1$, $i=1, \ldots, m$.
	\item[$2^\circ$] Set up a random order of $i$ going through $\{1,2,\ldots,m\}$. For each $i$, do Steps $3^\circ\sim 5^\circ$.
	\item[$3^\circ$] Denote 
	\[
	{\mathbf w}_i(z) = \left(\frac{1-z}{1-w_i} w_1, \ldots, \frac{1-z}{1-w_i} w_{i-1}, z, \frac{1-z}{1-w_i} w_{i+1}, \ldots, \frac{1-z}{1-w_i} w_m\right)^T
	\] 
	and $f_i(z) = f\{{\mathbf w}_i(z)\}$, $z\in [0,1]$.
	\item[$4^\circ$] Use an analytic solution or the quasi-Newton algorithm to find $z_*$ maximizing $f_i(z)$ with $z \in [0,1]$. Define ${\mathbf w}_*^{(i)} = {\mathbf w}_i(z_*)$. Note that $f({\mathbf w}_*^{(i)}) = f_i(z_*)$.
	\item[$5^\circ$] If $f({\mathbf w}_*^{(i)}) > f({\mathbf w}_a)$, replace ${\mathbf w}_a$ with ${\mathbf w}_*^{(i)}$, and $f\left({\mathbf w}_a\right)$ with $f({\mathbf w}_*^{(i)})$.
	\item[$6^\circ$] Repeat $2^\circ\sim 5^\circ$ until convergence, that is, $f({\mathbf w}_*^{(i)}) \leq f({\mathbf w}_a)$ for each $i$.
	\item[$7^\circ$] Report ${\mathbf w}_a$ as the D-optimal allocation.
\end{itemize}
\end{algorithm}

In practice, we may set up the stopping rule as $\frac{\max_{1\leq i\leq m} f(\mathbf w_*^{(i)}) }{\min_{1\leq i\leq m} f(\mathbf w_*^{(i)})} \le 1 + \epsilon$  for Step~$6^\circ$, where $\epsilon$ is a small positive number such as $10^{-8}$. Then Algorithm~\ref{algo:original_general_lift_one} guarantees a strict increase, which is at least $f({\mathbf w}_a)\cdot \epsilon$ in each round of iterations from Step~$2^\circ$ to Step~$5^\circ$. Since $S_0$ is compact, the maximum of $f({\mathbf w})$ on $S_0$ is finite. Algorithm~\ref{algo:original_general_lift_one} will stop in less than $\frac{\max_{{\mathbf w}\in S_0} f({\mathbf w}) - f({\mathbf w}_a)}{f({\mathbf w}_a) \epsilon}$ rounds of iterations. The same strategy can be used for Algorithm~1 as well.

\section{Commonly used GLM models}\label{sec:GLM_table}

In Table~\ref{tab:nuexamples}, we list commonly used GLM models, the corresponding link functions, and $\nu$ functions.

\begin{table}[ht]\caption{Examples of $\nu(\eta_i)$}\label{tab:nuexamples}
\begin{center}
\footnotesize
\begin{tabular}{|l|c|c|}\hline
Distribution of $Y_{ij}$ & Link function $g(\mu_i)$ & $\nu(\eta_i)$\\ \hline
Normal($\mu_i$, $\sigma^2$) & identity: $\mu_i$    & $\sigma^{-2}$ with known $\sigma^2>0$\\
Bernoulli($\mu_i$)    & logit: $\log \left(\frac{\mu_i}{1-\mu_i}\right)$             & $\frac{e^{\eta_i}}{(1+e^{\eta_i})^2}$\\
Bernoulli($\mu_i$)    & probit: $\Phi^{-1}(\mu_i)$            & $\frac{\phi^2(\eta_i)}{\Phi(\eta_i)\{1-\Phi(\eta_i)\}}$\\
Bernoulli($\mu_i$)    & c-log-log: $\log\{-\log(1-\mu_i)\}$         & $\frac{\exp(2\eta_i)}{\exp(e^{\eta_i})-1}$\\
Bernoulli($\mu_i$)    & log-log: $\log\{-\log(\mu_i)\}$           & $\frac{\exp(2\eta_i)}{\exp(e^{\eta_i})-1}$\\
Bernoulli($\mu_i$)    & cauchit: $\tan\{\pi\left(\mu_i - \frac{1}{2}\right)\}$           & $\frac{(1+\eta_i^2)^{-2}}{\pi^2/4 - \arctan^2(\eta_i)}$\\
Poisson($\mu_i$)      & log: $\log(\mu_i)$               & $\exp(\eta_i)$\\
Gamma($k$, $\mu_i/k$) & reciprocal: $\mu_i^{-1}$        & $k \eta_i^{-2}$ with known $k>0$\\
Inverse Gaussian($\mu_i$, $\lambda$) & inverse squared: $\mu_i^{-2}$    & $\lambda \eta_i^{-3/2}/4$ with known $\lambda>0$\\
\hline
\end{tabular}
\normalsize
\end{center}
\end{table}

\section{Two examples of finding $r_{i1}$ and $r_{i2}$ in Algorithm~1} \label{sec:two_ex_ri1_ri2}

Following the general procedure described in Subsection~3.2 for finding $r_{i1}$ and $r_{i2}$ in Step~$3^\circ$ of Algorithm~1, we provide the following two examples.

\begin{example}\label{ex:ni_Ni_n_r1_r2}{\rm 
If $S = \{(w_1, \ldots, w_m)^T \in S_0 \mid n w_i \leq N_i, i=1, \ldots, m\}$ as in Example~1, then ${\mathbf w}_i(z) \in S$ if and only if
\[
\left\{\begin{array}{c}
0\leq z \leq 1\\
n z \leq N_i\\
n \frac{1-z}{1-w_i}w_j \leq N_j, \> j\neq i
\end{array}
\right.
\]
which is equivalent to 
\[
\left\{\begin{array}{c}
0\leq z \leq 1\\
z \leq N_i/n\\
z \geq 1 - \frac{N_j}{n} \cdot \frac{1-w_i}{w_j}, \> j\neq i\mbox{ and }w_j>0
\end{array}
\right.
\]
Therefore, $r_{i1}$ and $r_{i2}$ in Step~$3^\circ$ of Algorithm~1 are 
\begin{equation}\label{eq:r1r2}
\left\{\begin{array}{cl}
r_{i1}  = & \max\left(\{0\} \cup \{1 - N_j/n \cdot (1-w_i)/w_j\mid j\neq i, w_j > 0\}\right)\\
r_{i2}  = & \min\{1, N_i/n\}
\end{array}\right.
\end{equation}
It can be verified that if ${\mathbf w} \in S$, then $0\leq r_{i1} \leq r_{i2}\leq 1$. 
\hfill{$\Box$}
}\end{example}

\begin{example}\label{ex:ninj_Nij_r1_r2}
If $S = \{(w_1, \ldots, w_m)^T \in S_0 \mid n \sum_{i=1}^4 w_i \leq 392, n \sum_{i=5}^8 w_i \leq 410\}$ as in Example~2, then $r_{i1}$ and $r_{i2}$ can be obtained as follows: 

\noindent
{\it Case one:} If $i \in \{1,2,3,4\}$, then ${\mathbf w}_i(z) \in S$ if and only if 
\begin{equation*}
\left\{\begin{array}{cl}
0 \leq z \leq 1 \\
z + \sum_{j=1, j\neq i}^4 \frac{w_j (1-z)}{1-w_i} \leq 392/n \\
\sum_{j=5}^8 \frac{w_j (1-z)}{1-w_i} \leq 410/n 
\end{array}\right.
\end{equation*}
which is equivalent to \begin{equation*}
\left\{\begin{array}{cl}
0 \leq z \leq 1 \\
z \leq \frac{392(1-w_i)-n\sum_{j=1,j \neq i }^4 w_j}{n(1-\sum_{j=1}^4 w_j)} \\
z \geq 1 - \frac{410(1-w_i)}{n\sum_{j=5}^8 w_j}
\end{array}\right.
\end{equation*}
Therefore, 
\begin{equation*}
 \left\{\begin{array}{cl}
 r_{i1} = & \max\{0, 1-\frac{410(1-w_i)}{n\sum_{j=5}^8 w_j}\}\\
r_{i2}  = & \min\{1, \frac{392(1-w_i)-n\sum_{j=1,j \neq i }^4 w_j}{n(1-\sum_{j=1}^4 w_j)}\}   
 \end{array}\right.
 \end{equation*}

\noindent
{\it Case two:} If $i \in \{5,6,7,8\}$, then ${\mathbf w}_i(z) \in S$ if and only if 
\begin{equation*}
\left\{\begin{array}{cl}
0 \leq z \leq 1 \\
\sum_{j=1}^4 \frac{w_j (1-z)}{1-w_i} \leq 392/n \\
z + \sum_{j=5, j\neq i}^8 \frac{w_j (1-z)}{1-w_i} \leq 410/n 
\end{array}\right.
\end{equation*}
Similarly, we obtain
\begin{equation*}
 \left\{\begin{array}{cl}
 r_{i1} = & \max\{0, 1 - \frac{392(1-w_i)}{n\sum_{j=1}^4 w_j} \}\\
r_{i2}  = & \min\{1, \frac{410(1-w_i)-n\sum_{j=5, j\neq i}^8 w_j}{n(1-\sum_{j=5}^8 w_j)}\}   
 \end{array}\right.
 \end{equation*}
\hfill{$\Box$}
\end{example}

\section{Another example of robustness under GLM models}\label{sec:another_GLM_robust}

\begin{example}\label{ex:GLM_modified_rubust}
To further test the robustness to model misspecifications in Example~6, we assume the true parameters to be  $\boldsymbol\beta = (\beta_0, \beta_1, \beta_{21},$ $\beta_{22} )^T  = (0,0.1,0.5,2)^T$, which is different from the one in Example~6. In this case, we have different D-optimal allocations for logit, probit, log-log, and complementary log-log links. Actually, using logit link, we obtain $\mathbf{w}_{\rm logit}=(0.189, 0.184, 0.050, 0.189, 0.181, 0.207)^T$. If the true link is probit with D-optimal allocation $\mathbf{w}_{\rm probit}=(0.193, 0.185, 0.050, 0.193, 0.181, 0.198)^T$, then the relative efficiency of ${\mathbf w}_{\rm logit}$ is $99.98\%$. Since the ${\mathbf W}$ matrix is the same for log-log and complementary log-log links, the corresponding D-optimal allocations are both $\mathbf{w}_{\rm log}=(0.189, 0.198, 0.050, 0.193, 0.198, 0.172)^T$. The relative efficiency of ${\mathbf w}_{\rm logit}$ with respect to ${\mathbf w}_{\rm log}$ is $99.68\%$. In other words, our D-optimal allocations are very robust with respect to link function misspecifications. 

We also provide in Table~\ref{tab:example2.1_model2_robustness} the average (sd) of RMSEs of the estimated coefficients from 100 independent simulations under Model~(4.4). The RMSE results in Table~\ref{tab:example2.1_model2_robustness} match the efficiency results. The logit link (the true link in the simulations) leads to the lowest average RMSE, while the other links have a little higher average RMSE. 
\end{example}

\begin{table}[ht] \caption{Average (sd) of RMSE of Estimated Parameters Based on Allocations Assuming Different Links over $100$ Simulations from Model~(4.4)}\label{tab:example2.1_model2_robustness}
\begin{adjustbox}{width={\textwidth},totalheight={\textheight},keepaspectratio}%
\begin{tabular}{ |c|c|c|c|c|c| }
 \hline
 \multirow{2}{4em}{Link Function}&\multicolumn{5}{c|}{Average (sd) of RMSE} \\\cline{2-6}
 & $\beta_0$ & all except $\beta_0$ & $\beta_1$ & $\beta_{21}$ & $\beta_{22}$\\
 \hline
 Logit & 0.231(0.166) & 1.123(0.609) & 0.269(0.198) & 0.268(0.177) & 0.423(0.370) \\
 Probit & 0.222(0.154)  &  1.142(0.620) & 0.258(0.212) & 0.276(0.228) & 0.452(0.340)  \\
 cloglog/loglog & 0.214(0.181) & 1.306(1.438) & 0.231(0.189)  &  0.261(0.190) & 0.779(2.249)  \\
 \hline
\end{tabular}
\end{adjustbox}
\end{table}

\section{Another example of trauma clinical study}\label{sec:another_trauma_example}

In this section, we provide an example where at least one constraint is attained at the D-optimal allocations.

\begin{example}\label{ex:modified_trauma}{\rm 
For the trauma clinical study described in Example~2, for illustration purposes, we consider the sampling problem with modified constraints as follows
\[
n(w_1+w_2+w_3+w_4) \leq 592,\>\>\> n(w_5+w_6+w_7+w_8) \leq 210
\]
with $n=600$.
In other words, we reduce the number of available severe cases to $210$. We derive allocations for different samplers as for Example~8, which are listed in Table ~\ref{tab:trauma_modified_allocation}. Note that the constraint $n(w_5+w_6+w_7+w_8) \leq 210$ is attained in both locally D-optimal and EW D-optimal allocations. It should also be noted that among the $B=1,000$ bootstrapped samples only $807$ fitted parameter vectors by SAS, in this case, are feasible. The quantiles of relative efficiencies of sampler allocations with respect to $807$ locally D-optimal allocations are listed in Table~\ref{tab:eff_trauma_modified}, which shows again that the EW D-optimal sampler is highly efficient with respect to the locally D-optimal allocations and much more efficient than the proportionally stratified and uniformly stratified samplers.  
\hfill{$\Box$}
}\end{example}

\begin{table}[ht] \caption{Allocations (Proportions) for Stratified Samplers in Example~\ref{ex:modified_trauma}}\label{tab:trauma_modified_allocation}
\begin{center}
\tiny
\begin{tabular}{ |c|c|c|c|c|c|c|c|c| }
 \hline
 \multicolumn{1}{|c|}{Severity} &\multicolumn{4}{c|}{Mild}&\multicolumn{4}{c|}{Severe} \\\hline
 \multicolumn{1}{|c|}{Dose} & 1 & 2 & 3 & 4 & 1 & 2 & 3 & 4\\\hline
 Proportional & \thead{116\\(0.193)} & \thead{105\\(0.175)} & \thead{115\\(0.192)} & \thead{108\\(0.180)} & \thead{41\\(0.068)} & \thead{37\\(0.062)} & \thead{40\\(0.067)} & \thead{38\\(0.063)} \\ \hline
  Uniform & \thead{98\\(0.163)} & \thead{98\\(0.163)} & \thead{97\\(0.162)} & \thead{97\\(0.162)} & \thead{55\\(0.092)} & \thead{50\\(0.083)} & \thead{54\\(0.090)} & \thead{51\\(0.085)}\\ \hline
  Locally D-opt ($\hat{\boldsymbol\theta}$) & \thead{234\\(0.390)} & \thead{4\\(0.007)} & \thead{3\\(0.005)} & \thead{149\\(0.249)} & \thead{126\\(0.210)} & \thead{0\\(0)} & \thead{3\\(0.005)} & \thead{81\\(0.134)}\\ \hline
  EW D-opt & \thead{253\\(0.421)} & \thead{0\\(0)} & \thead{0\\(0)} & \thead{137\\(0.229)} & \thead{77\\(0.128)} & \thead{8\\(0.013)} & \thead{0\\(0)} & \thead{125\\(0.209)}\\
 \hline
\end{tabular}
\normalsize
\end{center}
\end{table}

\begin{table}[ht]
\caption{Quantiles of Relative Efficiencies in Example~\ref{ex:modified_trauma}}\label{tab:eff_trauma_modified}
\begin{center}
\footnotesize
\begin{tabular}{ |c | c c c c c|}\hline
Sampler & Minimum & 1st Quartile & Median & 3rd Quartile & Maximum \\ \hline
SRSWOR & 51.80\% & 75.04\% & 75.80\% & 76.47\% & 78.98\% \\
Proportional & 52.36\% & 75.25\% & 75.62\% & 76.05\% & 77.42\% \\
Uniform & 57.19\% & 82.35\% & 82.71\% & 83.09\% & 84.29\% \\
EW D-opt & 69.05\% & 100\% & 100\% & 100\% & 100\% \\\hline
\end{tabular}
\normalsize
\end{center}
\end{table}

\section{Proofs}\label{sec:proofs}

\medskip\noindent
{\bf Proof of Lemma~1:}
If $w_i = 1$ for some $i$, then $w_j=0$ for all $j\neq i$ and $f({\mathbf w}) = \left|\sum_{j=1}^m w_j {\mathbf F}_j\right| = |{\mathbf F}_i| = 0$, which leads to a contradiction.
\hfill{$\Box$}

\medskip\noindent
{\bf Proof of Theorem~1:} Let ${\mathbf F}_i = (a_{ist})_{s,t=1,\ldots, p}$, $i=1, \ldots, m$. Then $\sum_{i=1}^m w_i {\mathbf F}_i = \left(\sum_{i=1}^m a_{ist} w_i\right)_{s,t=1,\ldots, p}$~. According to the definition of matrix determinant (see, for example, Section~4.4.1 in \cite{seber2008}), 
\[
f({\mathbf w}) = \left|\sum_{i=1}^m w_i {\mathbf F}_i \right| = \sum_{\pi} {\rm sgn}(\pi) \cdot \prod_{s=1}^p \left(\sum_{i=1}^m a_{is\; \pi(s)} w_i\right)
\]
is a homogeneous polynomial of $w_1, \ldots, w_m$, where $\pi$ goes through all permutations of $\{1,\ldots, p\}$, and ${\rm sgn}(\pi) = -1$ or $1$ depending on whether $\pi$ is odd or even. Since $f({\mathbf w}) > 0$ for some ${\mathbf w} \in S$, then $f({\mathbf w})$ is of order-$p$, not a zero function.

Since $f({\mathbf w}) = |\sum_{i=1}^m w_i {\mathbf F}_i|$ is a polynomial function of $w_1, \ldots, w_m$, then it must be continuous on $S$. According to the Weierstrass theorem (see, for example, Theorem~3.1 in \cite{sundaram1996first}), there must exist a ${\mathbf w}_* \in S$ such that $f({\mathbf w})$ attains its maximum at ${\mathbf w}_*$~.  
\hfill{$\Box$}

\medskip\noindent
\begin{lemma}\label{lem:positive-definite}
If $M_1, M_2 \in \mathbb{R}^{p\times p}$ are both positive semi-definite, then for any $\alpha \in (0,1)$, 
\[
 \log |\alpha M_1 + (1-\alpha) M_2| \ge \alpha \log |M_1| + (1-\alpha) \log |M_2|
\]
where the equality holds only if $M_1 = M_2$ or $|M_1|=|M_2|=0$.
\end{lemma}

\medskip\noindent
{\bf Proof of Lemma~\ref{lem:positive-definite}:}
When $M_1$ and $M_2$ are both positive definite, according to Theorem~1.1.14 in \cite{fedorov1972}, the inequality is always valid, and the equality holds only if $M_1 = M_2$~. If one of $M_1$ and $M_2$ is degenerate, then the right side of the equation $\alpha \log |M_1| + (1-\alpha) \log |M_2|= - \infty$. Since $ \log |\alpha M_1 + (1-\alpha) M_2| \ge - \infty$ is always true, the inequality is still valid when $M_1$  and $M_2$ are positive semi-definite matrices. If only one of $M_1$ and $M_2$ is degenerate, then $\alpha M_1 + (1-\alpha) M_2$ is still positive definite and only inequality holds.  
\hfill{$\Box$}

Lemma~\ref{lem:positive-definite} is an extended version of, for example, Theorem~1.1.14 in \cite{fedorov1972}. It is needed in the proof of  Lemma~2, which provides necessary results relevant to Step~$7^\circ$ of Algorithm~1.

\medskip\noindent
{\bf Proof of Lemma~2:} According to the constrained lift-one algorithm, ${\mathbf w}_a = (w_1, \ldots, $ $w_m)^T \in S$, $f({\mathbf w}_a) > 0$, and ${\mathbf w}_i(z) \in S$ for $z\in [r_{i1}, r_{i2}]$.

To avoid trivial cases, we assume $r_{i1} < r_{i2}$. For any $[z_1, z_2] \subseteq [r_{i1}, r_{i2}]$ and $\alpha \in (0,1)$, it can be verified that ${\mathbf w}_i\left\{\alpha z_1 + (1-\alpha) z_2\right\} = \alpha {\mathbf w}_i(z_1) + (1-\alpha) {\mathbf w}_i(z_2)$. Denote ${\mathbf w}_i(z_1) = (w_{11}, \ldots, w_{1m})^T \in S$ and ${\mathbf w}_i(z_2) = (w_{21}, \ldots,$ $ w_{2m})^T \in S$. According to Lemma~\ref{lem:positive-definite},
\begin{eqnarray*}
\log f_i\left\{\alpha z_1 + (1-\alpha) z_2\right\} &=& \log f\left[{\mathbf w}_i\left\{ \alpha z_1 + (1-\alpha) z_2\right\}\right]\\
&=& \log f\left\{\alpha {\mathbf w}_i(z_1) + (1-\alpha) {\mathbf w}_i(z_2)\right\}\\
&=& \log \left|\sum_{j=1}^m \left\{\alpha w_{1j} + (1-\alpha) w_{2j}\right\} {\mathbf F}_j\right|\\
&=& \log\left|\alpha \cdot \sum_{j=1}^m w_{1j} {\mathbf F}_j + (1-\alpha)\cdot \sum_{j=1}^m w_{2j} {\mathbf F}_j\right|\\
&\geq & \alpha \cdot \log \left|\sum_{j=1}^m w_{1j} {\mathbf F}_j\right| + (1-\alpha) \cdot \log \left|\sum_{j=1}^m w_{2j} {\mathbf F}_j\right|\\
&=& \alpha \log f\left\{{\mathbf w}_i(z_1)\right\} + (1-\alpha) \log f\left\{{\mathbf w}_i(z_2)\right\}\\
&=& \alpha \log f_i(z_1) + (1-\alpha) \log f_i(z_2)
\end{eqnarray*}
That is, $\log f_i(z)$ is a concave function on $[r_{i1}, r_{i2}]$.

If $z_*$ maximizes $f_i(z)$ with $z \in [r_{i1}, r_{i2}]$, then $f_i(z_*) \geq f_i(w_i) = f({\mathbf w}_a) > 0$. As a direct conclusion of  Theorem~1, $f_i(z)$ is a polynomial of $z$ and thus differentiable. Since $\log f_i(z)$ is concave, then $\partial \log f_i(z)/\partial z = f_i'(z)/f_i(z)$ is decreasing. The rest of the theorem is straightforward since  $f_i(z) > 0$ for all $z$ between $w_i$ and $z_*$~.
\hfill{$\Box$}

\medskip\noindent
{\bf Proof of Theorem~2:}
First of all, $f({\mathbf w}_*) \geq f({\mathbf w}) > 0$. Suppose ${\mathbf w}_*$ is not D-optimal in $S$. Then there exists a ${\mathbf w}_o = (w_1^o, \ldots, w_m^o)^T \in S$, such that, $f({\mathbf w}_o) > f({\mathbf w}_*) > 0$.

Denote ${\mathbf F}({\mathbf w}) = \sum_{i=1}^m w_i {\mathbf F}_i$ for ${\mathbf w} = (w_1, \ldots, w_m)^T$. Then ${\mathbf F}({\mathbf w})$ is a linear functional of ${\mathbf w}$, which implies ${\mathbf F}\{x {\mathbf w}_o + (1-x) {\mathbf w}_*\} = x {\mathbf F}({\mathbf w}_o) + (1-x) {\mathbf F}({\mathbf w}_*)$. Note that $f({\mathbf w}) = |{\mathbf F}({\mathbf w})|$. Since $f({\mathbf w}_o) > f({\mathbf w}_*) > 0$, then ${\mathbf F}({\mathbf w}_o) \neq {\mathbf F}({\mathbf w}_*)$ and $|{\mathbf F}({\mathbf w}_o)| > |{\mathbf F}({\mathbf w}_*)| > 0$. According to Lemma~\ref{lem:positive-definite}, $\log |{\mathbf F}\{x {\mathbf w}_o + (1-x) {\mathbf w}_*\}| = \log |x {\mathbf F}({\mathbf w}_o) + (1-x) {\mathbf F}({\mathbf w}_*)| > x \log |{\mathbf F} ({\mathbf w}_o)| + (1-x) \log |{\mathbf F} ({\mathbf w}_*)|$ for any $x \in (0,1)$.

We further denote ${\mathbf F}_x = {\mathbf F} \{x {\mathbf w}_o + (1-x) {\mathbf w}_*\}$ for $x\in [0,1]$. We claim that ${\mathbf F}_{x_1} \neq {\mathbf F}_{x_2}$ as long as $x_1 \neq x_2$~. Actually, if $x_1 \neq x_2$, then ${\mathbf F}_{x_1} = {\mathbf F}_{x_2}$ implies ${\mathbf F}({\mathbf w}_o) = {\mathbf F}({\mathbf w}_*)$, which is not true in this case. 

Now we define $f_*(x) = f\{x {\mathbf w}_o + (1-x) {\mathbf w}_*\} = |{\mathbf F} \{x {\mathbf w}_o + (1-x) {\mathbf w}_*\}| = |{\mathbf F}_x|$, $x\in [0,1]$. Then $\log f_*(x) = \log |{\mathbf F}\{x {\mathbf w}_o + (1-x) {\mathbf w}_*\}| > x \log |{\mathbf F}({\mathbf w}_o)| + (1-x) \log |{\mathbf F}({\mathbf w}_*)| > -\infty$ for each $x\in (0,1)$. Thus $f_*(x) > 0$ for each $x\in [0,1]$, which implies that the corresponding Fisher information matrix ${\mathbf F}_x$ is positive definite.

We claim that $\log f_*(x)$ is a strictly concave function on $x\in [0,1]$. Actually, for any $0\leq x_1 < x_2 \leq 1$ and any $\alpha \in (0,1)$, according to Lemma~\ref{lem:positive-definite},
\begin{eqnarray*}
& & \log f_* \{\alpha x_1 + (1-\alpha) x_2\}\\
&=& \log f\left[\alpha \{x_1 {\mathbf w}_o + (1-x_1) {\mathbf w}_*\} + (1-\alpha) \{ x_2 {\mathbf w}_o + (1-x_2) {\mathbf w}_*\}\right]\\
&=& \log |{\mathbf F}\left[\alpha \{x_1 {\mathbf w}_o + (1-x_1) {\mathbf w}_*\} + (1-\alpha) \{ x_2 {\mathbf w}_o + (1-x_2) {\mathbf w}_*\}\right]|\\
&=& \log |\alpha {\mathbf F}\left\{x_1 {\mathbf w}_o + (1-x_1) {\mathbf w}_*\right\} + (1-\alpha) {\mathbf F}\left\{ x_2 {\mathbf w}_o + (1-x_2) {\mathbf w}_*\right\}|\\
&=& \log |\alpha {\mathbf F}_{x_1} + (1-\alpha) {\mathbf F}_{x_2}|\\
&> & \alpha \log |{\mathbf F}_{x_1}|  + (1-\alpha)\log | {\mathbf F}_{x_2}|\\
&=& \alpha \log f_*(x_1) + (1-\alpha) f_*(x_2)
\end{eqnarray*}
As a direct conclusion, the first derivative of $\log f_*(x)$ is strictly decreasing as $x\in [0,1]$ increases. According to the mean value theorem (see, for example, Theorem~5.10 in \cite{rudin1976principles}), there exists a $c \in (0,1)$ such that
\begin{eqnarray*}
\left.\frac{\partial \log f_*(x)}{\partial x}\right|_{x=0} &\geq & \left.\frac{\partial \log f_*(x)}{\partial x}\right|_{x=c} = \frac{\log f_*(1)-\log f_*(0)}{1-0}\\
&=& \log f({\mathbf w}_o) - \log f({\mathbf w}_*) > 0
\end{eqnarray*}
Let $\varphi({\mathbf w}) = \log f({\mathbf w})$. Then the gradient of $\varphi({\mathbf w})$ is $\nabla \varphi({\mathbf w}) = f({\mathbf w})^{-1} \nabla f({\mathbf w})$. According to the definition of $f_*(x)$, the directional derivative of $f({\mathbf w})$ at ${\mathbf w}_*$ along ${\mathbf w}_o-{\mathbf w}_*$ is 
\begin{equation}\label{eq:wo_w*}
\nabla f({\mathbf w}_*)^T ({\mathbf w}_o-{\mathbf w}_*) = f({\mathbf w}_*) \cdot \nabla \varphi({\mathbf w}_*)^T ({\mathbf w}_o-{\mathbf w}_*) =  f({\mathbf w}_*) \cdot \left.\frac{\log f_*(x)}{\partial x}\right|_{x=0} > 0
\end{equation}

For $i=1, \ldots, m$, let $\bar{\mathbf w}_i = (0,\ldots, 0,1,0,\ldots, 0)^T \in \mathbb{R}^m$ whose $i$th coordinate is $1$. In the constrained lift-one algorithm at ${\mathbf w}_*$, we have ${\mathbf w}_i(z) = (1-\alpha) {\mathbf w}_* + \alpha \bar{\mathbf w}_i = {\mathbf w}_* + \alpha (\bar{\mathbf w}_i - {\mathbf w}_*)$ with $\alpha = \frac{z-w_i^*}{1-w_i^*}$. Note that $w_i^* < 1$ and ${\mathbf w}_i(w_i^*) = {\mathbf w}_*$~. It can be verified that the directional derivative of $f({\mathbf w})$ at ${\mathbf w}_*$ along $\bar{\mathbf w}_i-{\mathbf w}_*$ is
\begin{equation}\label{eq:directional_wi_w*}
\nabla f({\mathbf w}_*)^T (\bar{\mathbf w}_i-{\mathbf w}_*) =  (1-w_i^*) f_i'(w_i^*)
\end{equation}
Actually,
\begin{eqnarray*}
\nabla f({\mathbf w}_*)^T (\bar{\mathbf w}_i-{\mathbf w}_*) &=& f({\mathbf w}_*) \cdot \nabla \varphi({\mathbf w}_*)^T (\bar{\mathbf w}_i-{\mathbf w}_*)\\
&=& f({\mathbf w}_*) \cdot \lim_{\alpha\rightarrow 0} \frac{\varphi\left\{{\mathbf w}_* + \alpha (\bar{\mathbf w}_i - {\mathbf w}_*)\right\} - \varphi({\mathbf w}_*)}{\alpha}\\
\left(\mbox{replace $\alpha$ with $\frac{z-w_i^*}{1-w_i^*}$}\right) &=& f({\mathbf w}_*) (1-w_i^*)\cdot \lim_{z\rightarrow w_i^*} \frac{\varphi\left\{{\mathbf w}_i(z)\right\} - \varphi\{{\mathbf w}_i(w_i^*)\}}{z-w_i^*}\\
&=& f({\mathbf w}_*) (1-w_i^*)\cdot \lim_{z\rightarrow w_i^*} \frac{\log f\left\{{\mathbf w}_i(z)\right\} - \log f\{{\mathbf w}_i(w_i^*)\}}{z-w_i^*}\\
&=& f({\mathbf w}_*) (1-w_i^*)\cdot \lim_{z\rightarrow w_i^*} \frac{\log f_i(z) - \log f_i(w_i^*)}{z-w_i^*}\\
&=& f({\mathbf w}_*) (1-w_i^*)\cdot \left.\frac{\partial \log f_i(z)}{\partial z}\right|_{z=w_i^*}\\
&=& f({\mathbf w}_*) (1-w_i^*)\cdot \frac{f_i'(w_i^*)}{f_i(w_i^*)}\\
&=& (1-w_i^*) f_i'(w_i^*)
\end{eqnarray*}

Since $\sum_{i=1}^m w_i^o = 1$, then ${\mathbf w}_o - {\mathbf w}_* = \sum_{i=1}^m w_i^o(\bar{\mathbf w}_i-{\mathbf w}_*)$.  Since 
$f_i'(w_i^*)\leq 0$ for each $i$, 
then
\[
\nabla f({\mathbf w}_*)^T ({\mathbf w}_o - {\mathbf w}_*) = \sum_{i=1}^m w_i^o \nabla f({\mathbf w}_*)^T (\bar{\mathbf w}_i - {\mathbf w}_*) = \sum_{i=1}^m w_i^o (1-w_i^*) f_i'(w_i^*) \leq 0
\]
which leads to a contradiction with \eqref{eq:wo_w*}.
\hfill{$\Box$}

\medskip\noindent
{\bf Proof of Corollary~1:}
First of all, $f({\mathbf w}_*) >0$. Denote ${\mathbf w}_* = (w_1^*, \ldots,$ $ w_m^*)^T \in S_0$. Then $0\leq w_i^* < 1$ for each $i$ according to Lemma~1. 
Since $w_i^* < r_{i2}$ for each $i$, we have $f_i'(w_i^*) \leq 0$ according to Lemma~2.
Then ${\mathbf w}_*$ must be D-optimal in $S$ as a direct conclusion of Theorem~2.
\hfill{$\Box$}

\medskip\noindent
{\bf Proof of Theorem~3:}
There are two cases for ${\mathbf w}_*$ reaching Step~$10^\circ$. 

{\it Case one:} ${\mathbf w}_*$ is a converged allocation in Step~$6^\circ$ and satisfies the conditions in Step~$7^\circ$, that is, $f_i'(w_i^*) \leq 0$ for each $i$. According to Corollary~1, ${\mathbf w}_*$ must be D-optimal in $S$.

{\it Case two:} ${\mathbf w}_*$ is a converged allocation in Step~$6^\circ$, which satisfies the condition in Step~$8^\circ$ but violates some condition in Step~$7^\circ$. That is, $f_i'(w_i^*) > 0$ for some $i$ but $\max_{{\mathbf w} \in S} g({\mathbf w}) \leq 0$, where $g({\mathbf w}) = \sum_{i=1}^m w_i(1-w_i^*) f_i'(w_i^*)$. Since ${\rm rank}({\mathbf F}_i) < p$ for each $i$ and $f({\mathbf w}_*) > 0$, according to Lemma~1, $w_i^* < 1$ for each $i$.  Suppose ${\mathbf w}_*$ is not D-optimal in $S$. Then there exists a ${\mathbf w}_o = (w_1^o, \ldots, w_m^o)^T \in S$, such that, $f({\mathbf w}_o) > f({\mathbf w}_*) > 0$. According to the proof of Theorem~2,
\[
0 < \nabla f({\mathbf w}_*)^T ({\mathbf w}_o - {\mathbf w}_*) = \sum_{i=1}^m w_i^o (1-w_i^*) f_i'(w_i^*) = g({\mathbf w}_o)
\]
which contradicts the condition $\max_{{\mathbf w}\in S} g({\mathbf w}) \leq 0$ in Step~$8^\circ$. Therefore, ${\mathbf w}_*$ must be D-optimal in $S$.
\hfill{$\Box$}

\medskip\noindent
{\bf Proof of Theorem~5:}
(i) If $\sum_{i=1}^m c_i = 1$, then $S=\{(c_1, \ldots, c_m)^T\}$ which implies that ${\mathbf w}_o = (c_1, \ldots, c_m)^T$ is the only solution.

(ii) Suppose $\sum_{i=1}^m c_i > 1$. Without any loss of generality, we assume that $a_1 \geq a_2 \geq \cdots \geq a_m$. There exists a unique $k \in \{1, \ldots, m-1\}$ such that $\sum_{l=1}^k c_l \leq 1 < \sum_{l=1}^{k+1} c_l$. It can be verified that ${\mathbf w}_o = (c_1, \ldots, c_k, 1-\sum_{l=1}^k c_l, 0, \ldots, 0)^T$ maximizes $g({\mathbf w}) = \sum_{i=1}^m a_i w_i$. The rest part is straightforward.
\hfill{$\Box$}

\medskip\noindent
{\bf Proof of Lemma~3:} According to the proof of Theorem~1, 
\[
h(\alpha) = 
f\{(1-\alpha){\mathbf w}_*+\alpha {\mathbf w}_o\} = \sum_{\pi} {\rm sgn}(\pi) \cdot \prod_{s=1}^p \left[\sum_{i=1}^m a_{is\; \pi(s)} \{w_i^* + \alpha (w_i^o - w_i^*)\}\right]
\]
is an order-$p$ polynomial of $\alpha$. The rest of the lemma is straightforward.
\hfill{$\Box$}

\medskip\noindent
{\bf Proof of Theorem~6:} 
First of all, we claim that ${\mathbf F}({\mathbf w}_o) \neq {\mathbf F} ({\mathbf w}_*)$, where ${\mathbf F} ({\mathbf w}) = \sum_{i=1}^m w_i {\mathbf F}_i$ is the Fisher information matrix corresponding to the allocation ${\mathbf w} = (w_1, \ldots, $ $w_m)^T$. Actually, if ${\mathbf F}({\mathbf w}_o) = {\mathbf F} ({\mathbf w}_*)$, then $h(\alpha) = f\{(1-\alpha) {\mathbf w}_* + \alpha {\mathbf w}_o\} = |{\mathbf F}\{(1-\alpha) {\mathbf w}_* + \alpha {\mathbf w}_o\}| = |(1-\alpha) {\mathbf F}({\mathbf w}_*) + \alpha {\mathbf F} ({\mathbf w}_o)| \equiv |{\mathbf F}({\mathbf w}_*)|$. It implies $h'(0) = 0$. On the other hand, we denote $\varphi({\mathbf w}) = \log f({\mathbf w})$, then $\nabla\varphi({\mathbf w}) = f({\mathbf w})^{-1} \nabla f({\mathbf w})$ and 
\begin{eqnarray*}
g({\mathbf w}_o) &=& \nabla f({\mathbf w}_*)^T ({\mathbf w}_o - {\mathbf w}_*)\\
&=& f({\mathbf w}_*) \cdot \nabla \varphi({\mathbf w}_*)^T ({\mathbf w}_o - {\mathbf w}_*)\\
&=& f({\mathbf w}_*) \cdot \lim_{\alpha \rightarrow 0} \frac{\varphi \{{\mathbf w}_* + \alpha ({\mathbf w}_o - {\mathbf w}_*)\} - \varphi ({\mathbf w}_*)}{\alpha}\\
&=& f({\mathbf w}_*) \cdot \lim_{\alpha \rightarrow 0} \frac{\varphi \{(1-\alpha){\mathbf w}_* + \alpha {\mathbf w}_o\} - \varphi ({\mathbf w}_*)}{\alpha}\\
&=& f({\mathbf w}_*) \cdot \lim_{\alpha\rightarrow 0} \frac{\log h(\alpha) - \log h(0)}{\alpha}\\
&=& f({\mathbf w}_*) \cdot \frac{h'(0)}{h(0)}\\
&=& h'(0)
\end{eqnarray*}
Note that $h(0) = f({\mathbf w}_*) > 0$. Then $g({\mathbf w}_o) > 0$ implies $h'(0)>0$, which leads to a contradiction. We must have ${\mathbf F}({\mathbf w}_o) \neq {\mathbf F} ({\mathbf w}_*)$.

{\it (i)}  
Note that $h(\alpha) = f\left\{(1-\alpha) {\mathbf w}_* + \alpha {\mathbf w}_o\right\}$ is the same as the function $f_*(x)$ defined in the proof of Theorem~2. Since ${\mathbf F}({\mathbf w}_o) \neq {\mathbf F} ({\mathbf w}_*)$ and $|{\mathbf F} ({\mathbf w}_*)|>0$, we still have $h(\alpha) = f_*(\alpha) > 0$ for any $\alpha \in (0,1)$. Combining $h(0) = f({\mathbf w}_*) > 0$, we have $h(\alpha)  > 0$ for any $\alpha \in [0,1)$. Note that $h(1) = f({\mathbf w}_o)$ could be zero. 

{\it (ii)} $h'(0) > 0$ since $h'(0) = g({\mathbf w}_o) > 0$.

Since ${\mathbf F}({\mathbf w}_o) \neq {\mathbf F} ({\mathbf w}_*)$, we still have ${\mathbf F}_{x_1} \neq {\mathbf F}_{x_2}$ given $x_1 \neq x_2$ as in the proof of Theorem~2. Then $\log h(\alpha)$ is strictly concave for $\alpha \in [0,1)$ and $\frac{h'(\alpha)}{h(\alpha)}$ is strictly decreasing as $\alpha$ increases in $[0,1)$.

{\it (iii)} If $h(1) > 0$ and $h'(1) \geq 0$, then $\frac{h'(\alpha)}{h(\alpha)}$ is strictly decreasing as $\alpha$ increases in $[0,1]$. Since $\frac{h'(1)}{h(1)} \geq 0$, then $\frac{h'(\alpha)}{h(\alpha)} > \frac{h'(1)}{h(1)} \geq 0$ implies $h'(\alpha) > 0$ for all $\alpha \in (0,1)$. Therefore, $h(\alpha)$ attains its maximum at $\alpha_*=1$ only.

{\it (iv)} If $h(1) > 0$ and $h'(1) < 0$, then $\frac{h'(1)}{h(1)} < 0$. Since $\frac{h'(\alpha)}{h(\alpha)}$ is strictly decreasing on $\alpha \in [0,1]$, then there is one and only one $\alpha_* \in (0,1)$ such that $\frac{h'(\alpha_*)}{h(\alpha_*)} = 0$. That is, $h'(\alpha) > 0$ if $0\leq \alpha < \alpha_*$; $=0$ if $\alpha=\alpha_*$; and $<0$ if $\alpha_* < \alpha \leq 1$. Therefore, $h(\alpha)$ attains its maximum at $\alpha_* \in (0,1)$ only.

If $h(1) = f({\mathbf w}_o) =0$, we must have some $\alpha_{-} \in (0,1)$, such that $h'(\alpha_{-}) < 0$ since $h(0) > h(1)$. Since $\frac{h'(\alpha)}{h(\alpha)}$ is strictly decreasing on $\alpha \in [0,1)$, then there is one and only one $\alpha_* \in (0,\alpha_{-})$ such that $\frac{h'(\alpha_*)}{h(\alpha_*)} = 0$. That is, $h'(\alpha) > 0$ if $0\leq \alpha < \alpha_*$; $=0$ if $\alpha=\alpha_*$; and $<0$ if $\alpha_* < \alpha < 1$. Therefore, $h(\alpha)$ attains its maximum at $\alpha_* \in (0,1)$ only.

Since in general $h(1) = f({\mathbf w}_o) \geq 0$, cases~(iii) and (iv) actually cover all scenarios. Therefore, $\alpha_*$ exists and is unique all the time.
\hfill{$\Box$}

\medskip\noindent
{\bf Proof of Lemma~5:}
First of all, ${\mathbf w}_*$ exists and is unique. Actually, ${\mathbf w}_*$ exists since $S = \{{\mathbf w} \in S_0 \mid 0 \leq w_i \leq c_i, i=1, \ldots, m\}$ is bounded and closed. 

Secondly, ${\mathbf w}_*$ is unique and $f({\mathbf w}_*) > 0$. Actually, we denote $S_+ = \{{\mathbf w} \in S \mid f({\mathbf w}) > 0\}$, which is not empty since $\sum_{i=1}^m c_i \geq 1$. Given ${\mathbf w}_{(i)} = (w_1^{(i)}, \ldots, w_m^{(i)})^T \in S_+$, $i=1,2$, by letting $M_i = {\rm diag} \{w_1^{(i)}, \ldots, w_m^{(i)}\}$ in Lemma~\ref{lem:positive-definite}, it can be verified that $\log f\{\alpha {\mathbf w}_{(1)} + (1-\alpha) {\mathbf w}_{(2)}\} > \alpha \log f({\mathbf w}_{(1)}) + (1-\alpha) \log f({\mathbf w}_{(2)})$ for all $\alpha \in (0,1)$ if ${\mathbf w}_{(1)} \neq {\mathbf w}_{(2)}$~. In other words, $\log f({\mathbf w})$ is strictly concave on $S_+$, which leads to the uniqueness of ${\mathbf w}_*$~.

{\it Case (i):} If without the constraints $w_i \leq c_i$, ${\mathbf w}_* = (1/m, \ldots, 1/m)^T$ maximizes $f({\mathbf w})$ due to the relationship between geometric average and arithmetic average. If $\min_{1\leq i\leq m} c_i \geq 1/m$, then such a ${\mathbf w}_*$ belongs to $S$ and thus is also the solution with constraints.

{\it Case (ii):} Without any loss of generality, we assume $c_1 \leq \cdots \leq c_m$~. Then $c_i = c_{(i)}$, $i=1, \ldots, m$. Similarly, we let $c_{m+1}=1$. Note that $c_1 = \min_{1\leq i\leq m} c_i < 1/m$ and $c_m = \max_{1\leq i\leq m} c_i \leq 1$.

First we show that there exist $k\in \{1, \ldots, m-1\}$ and $u \in [c_k, c_{k+1})$ that ${\mathbf w}_* := (c_1, \ldots, c_k, u,$ $\ldots, u)^T \in S$, that is, $\sum_{i=1}^k c_i + (m-k)u = 1$. Actually, if we define
\[
h(x) = \left\{\begin{array}{cl}
mx & \mbox{ if }0\leq x< c_1\\
\sum_{i=1}^l c_i + (m-l)x & \mbox{ if }c_l \leq x < c_{l+1}, l =1, \ldots, m-1\\
\sum_{i=1}^m c_i & \mbox{ if }x\geq c_m
\end{array}\right.
\]
then $h(x)$ is continuous on $[0,1]$ and is strictly increasing on $[0, c_m]$. Since $h(0)=0$ and $h(c_m) = \sum_{i=1}^m c_i > 1$, then there exist a unique $u \in (0, c_m) = (0, \max_{1\leq i\leq m} c_i)$ and a corresponding $1\leq k\leq m-1$ such that $h(u) = \sum_{i=1}^k c_i + (m-k)u = 1$. 

Secondly, we show that ${\mathbf w}_* = (c_1, \ldots, c_k, u, \cdots, u)^T$ is a converged allocation in Step~$6^\circ$ of Algorithm~1. Actually, for $1\leq i\leq k$, $w_i = c_i$, $r_{i1} = r_{i2} = c_i$ for Step~$3^\circ$ of Algorithm~1, which leads to $z_* = c_i$~. Note that in this case, $f'_i(z) = c_i^{-1} \prod_{l=1}^kc_l u^{m-k} (1-c_i)^{1-m} (1-z)^{m-2} (1-mz)$ and thus $f'_1(z_*) = f'_1(c_1) > 0$. For $k+1 \leq i \leq m$, $w_i = u < c_i$, $r_{i1} = u$ and $r_{i2} = c_i$, $f'_i(z) = \prod_{i=1}^k c_i u^{m-k-1} (1-u)^{1-m} (1-z)^{m-2} (1-mz) < 0$ for all $z\in [u, c_i]$, which leads to $z_* = u$ in this case.

Thirdly, we show that $\max_{{\mathbf w} \in S} g({\mathbf w}) = 0$ as defined in Step~$8^\circ$ in Algorithm~1. It can be verified that in this case, for ${\mathbf w} = (w_1, \ldots, w_m)^T \in S$
\[
g({\mathbf w}) = \prod_{l=1}^k c_l \cdot u^{m-k} \left( \sum_{i=1}^k  c_i^{-1} w_i + u^{-1} \sum_{i=k+1}^m w_i - m \right) 
\]
Since $c_1^{-1} \geq c_2^{-1} \geq \cdots \geq c_k^{-1} \geq u^{-1} > 0$, it can be verified that ${\mathbf w}_*$ also maximizes $g({\mathbf w})$ and $g({\mathbf w}_*) = 0$.

By applying Theorem~3 to GLMs with $m=p$, it can be verified that ${\mathbf w}_*$ maximizes $f({\mathbf w})$ with ${\mathbf w} \in S$.

{\it Case (iii):} If $\sum_{i=1}^m c_i = 1$, then $S = \{(c_1, \ldots, c_m)^T\}$ and ${\mathbf w}_* = (c_1, \ldots,$ $c_m)^T$ is the only feasible solution.
\hfill{$\Box$}

\medskip\noindent
{\bf Proof of Theorem~7:}
For GLM~(4.1), if $m=p$, then $f({\mathbf w}) = |{\mathbf X}^T {\mathbf W} {\mathbf X}| = |{\mathbf X}|^2 \prod_{i=1}^m \nu_i \cdot \prod_{i=1}^m w_i$. According to Lemma~5, the constrained uniform allocation ${\mathbf w}_*$ maximizes $\prod_{i=1}^m w_i$, ${\mathbf w} \in S$. That is, ${\mathbf w}_*$ is D-optimal on $S$.

Similarly, since $f_{\rm EW} ({\mathbf w}) = |{\mathbf X}^T E({\mathbf W}) {\mathbf X}| = |{\mathbf X}|^2 \prod_{i=1}^m E(\nu_i) \cdot \prod_{i=1}^m w_i$, ${\mathbf w}_*$ is EW D-optimal on $S$ as well.
\hfill{$\Box$}

\end{document}